\documentclass[aps,pra,reprint,superscriptaddress,showpacs,showkeys]{revtex4-1}
\usepackage{amssymb}
\usepackage{amsmath}
\usepackage{cancel}
\usepackage{color}
\usepackage{braket}
\usepackage{graphicx}
\usepackage{epstopdf}
\usepackage{verbatim}
\usepackage{siunitx}
\usepackage{soul}
\usepackage{xr}
\usepackage[toc,page]{appendix}
\usepackage[section]{placeins}

\usepackage[unicode]{hyperref}
\usepackage{tikz}

\newcommand{\cnuc}{$^{\mathrm{13}}$C}

\renewcommand{\check}[1]{\textcolor{red}{#1}}

\newcommand*\circled[1]{\tikz[baseline=(char.base)]{
            \node[shape=circle,draw,inner sep=1pt] (char) {#1};}}

\begin{document}

\title{
An integrated nanophotonic quantum register based on silicon-vacancy spins in diamond
}

\author{C. T. Nguyen}
\email[These authors contributed equally to this work\\]{christiannguyen@g.harvard.edu}
\affiliation{Department of Physics, Harvard University, Cambridge, Massachusetts 02138, USA}

\author{D. D. Sukachev}
\email[These authors contributed equally to this work\\]{christiannguyen@g.harvard.edu}
\affiliation{Department of Physics, Harvard University, Cambridge, Massachusetts 02138, USA}

\author{M. K. Bhaskar}
\email[These authors contributed equally to this work\\]{christiannguyen@g.harvard.edu}
\affiliation{Department of Physics, Harvard University, Cambridge, Massachusetts 02138, USA}

\author{B. Machielse}
\email[These authors contributed equally to this work\\]{christiannguyen@g.harvard.edu}
\affiliation{Department of Physics, Harvard University, Cambridge, Massachusetts 02138, USA}
\affiliation{John A. Paulson School of Engineering and Applied Sciences, Harvard University, Cambridge, Massachusetts 02138, USA}

\author{D. S. Levonian}
\email[These authors contributed equally to this work\\]{christiannguyen@g.harvard.edu}
\affiliation{Department of Physics, Harvard University, Cambridge, Massachusetts 02138, USA}

\author{E. N. Knall}
\affiliation{John A. Paulson School of Engineering and Applied Sciences, Harvard University, Cambridge, Massachusetts 02138, USA}

\author{P. Stroganov}
\affiliation{Department of Physics, Harvard University, Cambridge, Massachusetts 02138, USA}

\author{C. Chia}
\affiliation{John A. Paulson School of Engineering and Applied Sciences, Harvard University, Cambridge, Massachusetts 02138, USA}

\author{M. J. Burek}
\affiliation{John A. Paulson School of Engineering and Applied Sciences, Harvard University, Cambridge, Massachusetts 02138, USA}

\author{R. Riedinger}
\affiliation{Department of Physics, Harvard University, Cambridge, Massachusetts 02138, USA}

\author{H. Park}
\affiliation{Department of Physics, Harvard University, Cambridge, Massachusetts 02138, USA}
\affiliation{Department of Chemistry and Chemical Biology, Harvard University, Cambridge, Massachusetts 02138, USA}

\author{M. Lon\v{c}ar}
\affiliation{John A. Paulson School of Engineering and Applied Sciences, Harvard University, Cambridge, Massachusetts 02138, USA}

\author{M. D. Lukin}
\email{lukin@physics.harvard.edu}
\affiliation{Department of Physics, Harvard University, Cambridge, Massachusetts 02138, USA}

\begin{abstract}
We realize an elementary quantum network node consisting of a silicon-vacancy (SiV) color center inside a diamond nanocavity coupled to a nearby nuclear spin with \SI{100}{ms} long coherence times. 
Specifically, we describe experimental techniques and discuss effects of strain, magnetic field, microwave driving, and spin bath on the properties of this 2-qubit register. 
We then employ these techniques to generate Bell-states between the SiV spin and an incident photon as well as between the SiV spin and a nearby nuclear spin. We also discuss control techniques and parameter regimes for utilizing the SiV-nanocavity system as an integrated quantum network node.
\end{abstract}

\maketitle

\section{Introduction}
\label{sec:intro}

    \begin{figure}[b]
		\includegraphics[width=\linewidth]{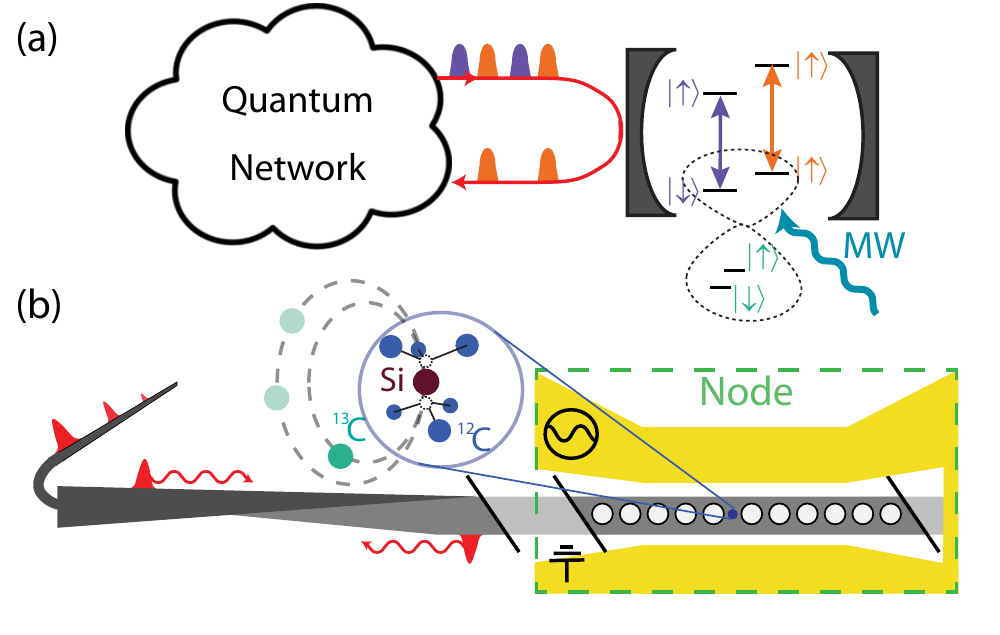}
	\caption{
			(a) Schematic of a quantum network. Nodes consisting of several qubits are coupled together via an optical interface.
			(b) A quantum network node based on the SiV. SiV centers and ancilla \cnuc\ are incorporated into a nanophotonic device and addressed with a coupled fiber and microwave coplanar waveguide.
			}
	    \label{fig:intro}
    \end{figure}
		
Quantum networks have the potential to enable a plethora of new technologies including secure communication, enhanced metrology, and distributed quantum computing \cite{kimble2008quantum,childress2005fault,gottesman2012sensing,komar2014quantum,monroe2014large}. 
Such networks require nodes which perform quantum processing on a small register of interconnected qubits with long coherence times. Distant nodes are connected by efficiently interfacing qubits with optical photons that can be coupled into an optical fiber [Fig.~\ref{fig:intro}(a)]. 

The prevailing strategy for engineering an efficient, coherent optical interface is that of cavity quantum electrodynamics (QED), which enhances the interactions between atomic quantum memories and photons \cite{stute2013quantum,reiserer2015cavity,kalb2015heralded,lodahl2015interfacing,welte2018photon}. 
Nanophotonic cavity QED systems are particularly appealing, as the tight confinement of light inside optical nanostructures enables strong, high-bandwidth qubit-photon interactions \cite{kim2011strong, Javadi2015, sipahigil2016integrated}. 
In practice, nanophotonic devices also have a number of technological advantages over macroscopic optical cavities, as they can be fabricated en-masse and interfaced with on-chip electronics and photonics, making them suitable for scaling up to large-scale networks \cite{lodahl2015interfacing, Molesky2018inverse}. 
While strong interactions between single qubits and optical photons have been demonstrated in a number of cavity QED platforms \cite{lodahl2015interfacing, sun2016quantum, abobeih2018one, Waldherr2014Quantum, Pfaff2014unconditional, welte2018photon}, no single realization currently meets all of the requirements of a quantum network node. 
Simultaneously achieving high-fidelity, coherent control of multiple long-lived qubits inside of a photonic structure is a major outstanding challenge.

Recent work has established the silicon-vacancy color-center in diamond (SiV) as a promising candidate for quantum networking applications \cite{becker2016ultrafast, sukachev2017silicon, becker2018all, evans2018photon, meesala2018strain, metsch2019initialization}. 
The SiV is an optically active point defect in the diamond lattice \cite{hepp2014electronic,muller2014optical}. 
Its $D_{3d}$ inversion symmetry results in a vanishing permanent electric dipole moment of the ground and excited states, rendering the transition insensitive to electric field noise typically present in nanostructures \cite{evans2016narrow}. 
Recent work has independently shown that SiV centers in nanostructures display strong interactions with single photons \cite{evans2018photon} and that SiV centers at temperatures below \SI{100}{\milli\kelvin} (achievable in dilution refrigerators) exhibit long coherence times \cite{jahnke2015electron,sukachev2017silicon}. 
While these results indicate the promising potential of the SiV center for future quantum network nodes, significant technical challenges must be overcome in order to combine these ingredients.

In this paper, we outline the practical considerations and approaches needed to build a quantum network node with SiV centers in nanophotonic diamond cavities coupled to ancillary nuclear spins [Fig.~\ref{fig:intro}(b)] \cite{nguyen2019quantum}. 
Section \ref{sec:fab} describes recent improvements to the fabrication techniques used to create and incorporate SiV centers into high-quality factor, critically-coupled nanophotonic cavities with an efficient fiber-optical interface.
Section \ref{sec:setup} describes the millikelvin experimental apparatus and several common experimental protocols. 
Section \ref{sec:strain} describes the SiV level structure and electronic transitions, illusutrating the interplay of strain and magnetic field in enabling both coherent control of\textendash\ and a photonic interface for\textendash\ SiV spins.
Sections \ref{sec:cavity}, \ref{sec:mwcontrol} and \ref{sec:mwdeer} outline experimental implementations of optical and microwave control of SiV centers, and use this control to create electron-photon Bell states with high fidelity in section \ref{sec:spinphoton}. 
Section \ref{sec:Ncontrol} introduces techniques for coupling to additional qubits consisting of naturally occuring \cnuc\ in diamond. We describe our method for initializing and reading out these nuclear spins via the SiV, coherent control of \cnuc\ with microwave and radio-frequency driving, probe the coherence of these nuclei, and finally entangle the SiV with a nearby \cnuc\ and demonstrate electron-nuclear Bell states. 

\section{Nanophotonic device fabrication}
\label{sec:fab}

    \begin{figure}
		\includegraphics[width=\linewidth]{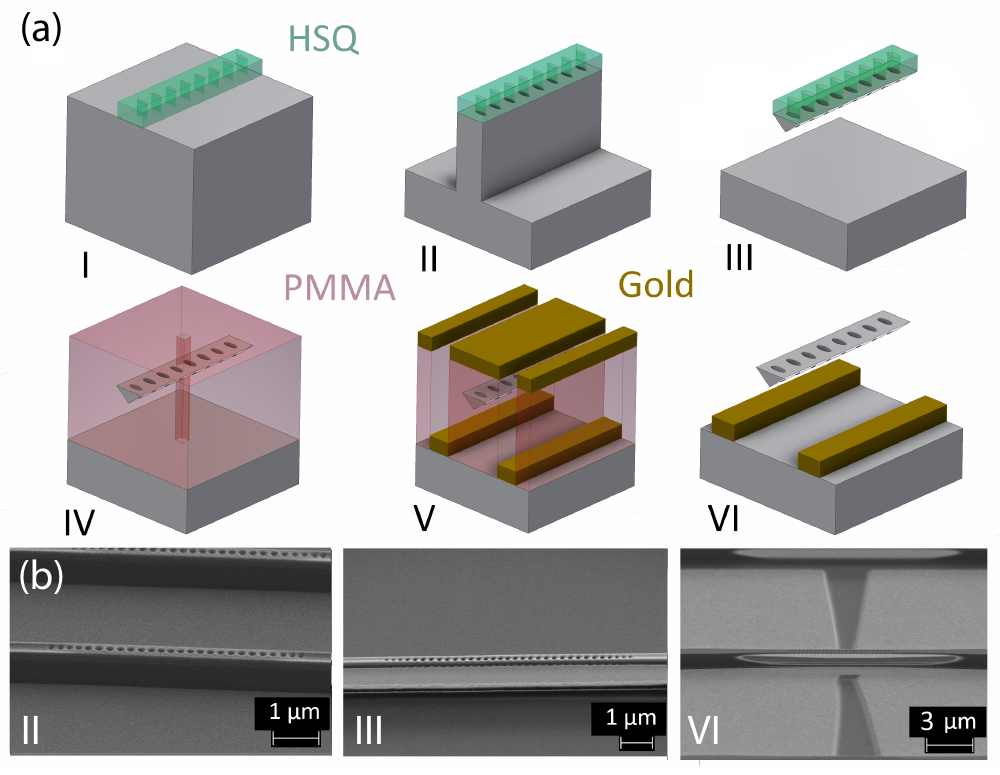}
	\caption{
			(a) Schematic of the nanofabricataion process used to produce devices. I: Titanium-HSQ mask is patterned using EBL. II: Pattern is transferred onto diamond using top down $\mathrm{O}_2$ RIE. III: Angled IBE is used to separate structures from substrate. IV: Devices are covered in PMMA and implantation aperatures are formed using EBL. Device are then cleaned, implanted, and annealed. V: PMMA is used in a litoff procedure to pattern gold microwave striplines. VI: Final devices are cleaned and prepared for experiment.
			(b) Scanning electron micrographs corresponding to steps II, III, and VI in the fabricaton procedure.
			}
	    \label{fig:devices}
    \end{figure}
		
\subsection{Device design}
The devices used in these experiments integrate nanophotonic cavities, implanted SiV centers, and microwave coplanar waveguides onto a single diamond chip. 
Here we present the fabrication process used to realize such devices. 

Typically, high-quality photonic crystal resonators are fabricated from 2-D membranes, which tightly confine light due to total internal reflection off of material boundaries.
Difficulties in growing high-purity, single-crystal diamond films on non-diamond substrates are one of the key challenges to fabricating such resonators in diamond \cite{bauer1990fundamental}.
As a result, nanophotonic diamond structures must be etched out of bulk diamond, which requires non-traditional etching techniques \cite{burek2012free, Khanaliloo2015high}. 
In particular, two methods have emerged for creating freestanding diamond nanostructures: Isotropic undercutting \cite{Khanaliloo2015high, Mouradian2017rectangular} and angled ion-beam etching (IBE) \cite{Atikian2017freestanding}. 
In this work, we use the latter technique, resulting in freestanding, triangular-cross-section waveguides.

Preliminary design of the nanophotonic structures are described in appendix \ref{apx:fab}, and are optimized to maximize atom-photon interaction while maintaining high waveguide coupling. 
To take advantage of the scalable nature of nanofabrication, these optimized devices are patterned in sets of roughly 100 with slightly modified fabrication parameters. 
The overall scale of all photonic crystal cavity parameters are varied between different devices on the same diamond chip to compensate for fabrication errors (which lead to unexpected variations in the resonator frequency and quality-factor). 
Due to these errors, roughly one in six cavities are suitable for SiV experiments. Fortunately, hundreds of devices are made in a single fabrication run, ensuring that every run yields many usable devices. 

The diamond waveguide region (as opposed to the photonic crystal cavity region [Appendix.~\ref{apx:fab}]) has two distinguishing features. 
First, thin support structures are placed periodically along the waveguide and are used to suspend the structures above the substrate. 
These supports are portions of the waveguide which are adiabatically tapered to be $\sim 30\%$ wider than the rest of the waveguide, and take longer to etch away during the angled etch process. 
By terminating the etch after normal waveguide regions are fully etched through, these wide sections become $\sim$ \SI{10}{\nano\meter} thick supports which tether the waveguide structures to the substrate while minimizing scattered loss from guided modes. 
Second, one end of the waveguide structure is adiabatically tapered into free-space \cite{burek2017fiber}. 
These tapers are formed by a linear taper of the waveguide down to less than \SI{50}{\nano\meter} wide over a \SI{10}{\micro\meter} length. 
This tapered region can be coupled to a similarly tapered optical fiber, allowing structures to efficiently interface with a fiber network [Sec.~\ref{sec:setup}].
This tapered end of the waveguide is the most fragile portion of the structure, and can break after repeated fiber coupling attempts. This is often what limits the total measurement lifetime of a device. 

The number of devices (and thus the relative yield of the fabrication process) is limited by the maximum packing density on the diamond chip. 
This is primarily limited by the need to accommodate \SI{10}{\micro\meter} wide microwave coplanar waveguides (CPWs) between devices, which are patterned directly onto the diamond surface to efficiently control SiV spins using microwaves. 
Simulations (Sonnet Inc) of prospective design geometries ensure that the CPW is impedance matched with our \SI{50}{\ohm} feed lines, which minimizes scattered power from the waveguides. 
Tapers in the CPW near the center of the cavity regions concentrate current and increase the amplitude of the microwave field near the SiVs, and CPWs are terminated with a short in order to ensure a magnetic field maximum along the device. 

\subsection{Device fabrication}
Fabrication of the diamond structures proceeds as described in ref.~\cite{burek2017fiber} with the notable modification that the angled etch is conducted not with a Faraday cage loaded inside a reactive ion etching chamber, but instead with an IBE. 
The Faraday cage technique \cite{burek2012free, burek2014high} offered the benefit of simplicity and accessibility\textemdash requiring only that the reactive ion etching chamber in question was large enough to accommodate the cage structure\textemdash but suffered from large fluctuations in etch rate across the surface of the sample, as well as between different fabrication runs, due to imperfections in the Faraday cage mesh. 
These irregularities could be partially compensated for by repeatedly repositioning and rotating the cage with respect to sample during the etch, but this process proved to be laborious and imprecise. 
Instead, IBE offers collimated beams of ions several \si{\centi\meter} in diameter, leading to almost uniform etch rates across the several \si{\milli\meter} diamond chip. This technique allowed for consistent fabrication of cavities with $Q>10^4$, $V<0.6 [\lambda/(n=2.4)]^3$, and resonances within $\sim$ \SI{10}{\nano\meter} of SiV optical frequencies.

Once the diamond cavities are fabricated [Fig.~\ref{fig:devices}(a I-III)], SiV centers must be incorporated. 
To ensure the best possible atom-photon interaction rate [Sec.~\ref{sec:cavity}], SiVs should be positioned at the cavity mode maximum. 
Ideally, this requires implantation accuracy of better than \SI{50}{\nano\meter} in all 3 dimensions due to the small mode volume ($\sim 0.5 [\lambda/(n=2.4)]^3$) of the cavities used. 
In the past, implantation of silicon ions (which form SiV centers following a high-temperature anneal) was done using focused ion-beam implantation, but this technique required specialized tools and lacked the accuracy necessary for maximally efficient mode coupling \cite{sipahigil2016integrated}. 
Instead, we adapt the standard masked implantation technique and use commercial foundaries for ion implantation.

For the implantation process, we repeatedly spin and bake MMA EL11 and PMMA C4 (Microchem) to cover the nanophotonic cavities completely with polymer resist. We then spin-coat a conductive surface layer of Espacer (Showa Denko). 
An E-beam lithography (EBL) tool then aligns with large markers underneath the polymer layer, allowing it to expose an area surrounding smaller, high-resolution alignment markers on the diamond. 
The exposed regions are developed in a 1:3 mixture of MIBK:IPA. Espacer is again spin-coated, and a second EBL write can be done, aligned to the high-resolution markers. 
Based on these alignment markers, holes of less than \SI{65}{\nano\meter} diameter (limited by the resolution of PMMA resist) are patterned onto the center of the photonic crystal cavity which, after subsequent development, act as narrow apertures to the diamond surface [Fig.~\ref{fig:devices}(a IV)]. 
The rest of the diamond surface is still covered in sufficiently thick PMMA to prevent ions from reaching masked portions of the device. 
Diamonds are then sent to a commercial foundry (Innovion) where they are implanted with silicon ions at the appropriate energy and dose [Fig.~\ref{fig:devices} (b)]. 
Annealing in a UHV vacuum furnace (Kurt-Lesker) at $\sim$\SI{1400}{\kelvin} converts these implanted ions into SiV centers \cite{clark1995silicon, evans2016narrow}.

CPWs are fabricated using a liftoff process similar to that used to create masked implantation windows. 
The most notable difference is an additional oxygen plasma descum after development to remove PMMA residue from the surface. 
Following development, a \SI{10}{\nano\meter} titanium film serves as an adhesion layer for a \SI{250}{\nano\meter} thick gold CPW [Fig.~\ref{fig:devices} (a V)]. 
Liftoff is performed in heated Remover PG (Microchem) [Fig.~\ref{fig:devices} (a VI)].
The metal thicknesses used here are chosen to improve adhesion of the gold, as well as prevent absorption of cavity photons by the metallic CPW. 
We observe that the cavity quality factor significantly degrades with gold films $>$ \SI{300}{\nano\meter}.
Due to ohmic heating, which can degrade the coherence properties of SiV spins [Sec.~\ref{sec:mwcontrol}], the length of the CPW is constrained to address a maximum of roughly 6 devices.

Future improvements in diamond device performance will be predicated on improvements of the fabrication technology. 
Device quality factors are currently limited by deviations in device cross section caused by imperfect selectivity of the HSQ hard mask to oxygen etching. 
Replacing this mask with a sufficiently smooth metal mask could result in improved etch selectivity and device performance. 
Isotropic undercut etching could also lead to improved control over device cross sections and facilitate more sophisticated device geometries \cite{Mouradian2017rectangular, Dory2019optimized} at the cost of reduced control over isotropically etched surface roughness. 
Various techniques exist for the formation of smaller implantation apertures \cite{Toyli2010chip, Staudacher2012enhancing}, but these techniques are difficult to use in conjunction with implantation into completed nanophotonic devices. 
Finally, the use of superconducting striplines could reduce heating, which would enable the CPW to potentially address all devices on the diamond chip and allow for faster driving of SiV spin and nuclear transitions [Sec.~\ref{sec:mwcontrol}, \ref{sec:Ncontrol}]. 

\section{Experimental Setup}
\label{sec:setup}

    \begin{figure}
		\includegraphics[width=\linewidth]{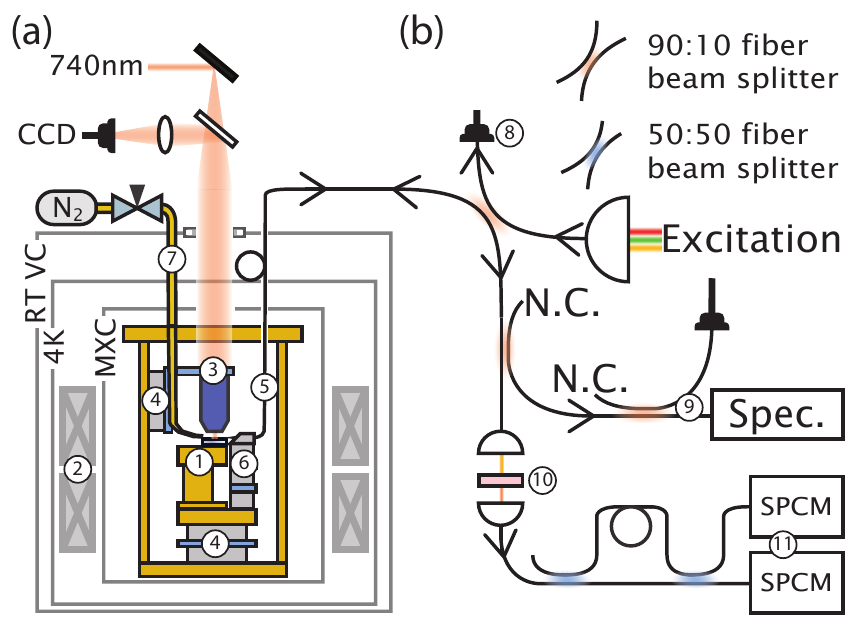}
	\caption{
			(a) Experiment schematic. Devices \protect\circled{\scriptsize{1}} are mounted in the bore of a SC magnet \protect\circled{\scriptsize{2}} inside of a dilution refrigerator, and imaged with wide-field imaging \protect\circled{\scriptsize{3}} and piezo steppers \protect\circled{\scriptsize{4}}. Devices are addressed with a tapered optical fiber \protect\circled{\scriptsize{5}} positioned using a second set of piezo steppers \protect\circled{\scriptsize{6}}. Cavities are tuned using a nitrogen \protect\circled{\scriptsize{7}}.
			(b) Fiber network used to probe devices. Excitation light is monitored \protect\circled{\scriptsize{8}} and sent to the device. Collected light is monitored \protect\circled{\scriptsize{9}} and filtered \protect\circled{\tiny{10}} then sent to one or several SPCMs \protect\circled{\tiny{11}}. N.C. indicates no connection.
			}
\label{fig:setup}
    \end{figure}

Experiments are performed in a home-built photonic-probe setup inside of a dilution refrigerator (DR, BlueFors BF-LD250) [Fig.~\ref{fig:setup}(a)]. 
The diamond substrate is mounted to a gold-plated copper sample holder via indium soldering below the mixing chamber in the bore of a (6,1,1) \si{\tesla} superconducting vector magnet (American Magnetics Inc.) anchored to the \SI{4}{\kelvin} stage. 
A thermal link between the device and the mixing chamber plate is provided by gold-plated copper bars, as well as oxygen-free copper braids (Copper Braid Products), ensuring maximal thermal conductivity between the mixing chamber plate and the sample, which reaches a base temperature of roughly \SI{60}{\milli\kelvin}. 
We address single nanophotonic devices via a tapered optical fiber, which can be coupled \textit{in-situ} with collection efficiencies exceeding $90\%$ \cite{burek2017fiber}. 
The tapered fiber is mounted to a 3-axis piezo stepper (ANPx101, ANPz101), and imaged in free-space by an $8f$ wide-field scanning confocal microscope which focuses onto a cryo-compatible objective (Attocube LT-APO-VISIR). 
This setup allows for coupling to several cavities during a single cooldown. 

Once coupled, the cavity resonance is red-shifted via nitrogen gas condensation \cite{evans2018photon}. 
A copper tube is weakly thermalized with the \SI{4}{\kelvin} plate of the DR and can be heated above \SI{80}{\kelvin} in order to flow $N_2$ gas onto the devices. 
This gas condenses onto the photonic crystal, modifying its refractive index and red-shifting the cavity resonance. 
When the copper tube is not heated, it thermalizes to \SI{4}{\kelvin}, reducing the blackbody load on the sample and preventing undesired gas from leaking into the vacuum chamber. 

After red-tuning all devices in this way, each cavity can be individually blue-tuned by illuminating the device with a $\sim$\SI{100}{\micro\watt} broadband laser via the tapered fiber, locally heating the device and evaporating nitrogen. 
This laser-tuning can be performed very slowly to set the cavity resonance with a few \si{\giga\hertz}.
The cavity tuning range exceeds \SI{10}{\nano\meter} without significantly degrading the cavity quality factor, and is remarkably stable inside the DR, with no observable drift over several months of measurements.

In previous work \cite{evans2018photon}, SiVs were probed in transmission via the free-space confocal microscope focused onto a notch opposing the tapered fiber. 
Mechanical vibrations arising from the DR pulse tube ($\sim$\SI{1}{\micro\meter} pointing error at the sample position) result in significant fluctuations in power and polarization of incoupled light. 
In this work, we demonstrate a fully integrated solution by utilizing the same tapered fiber to both probe the device and collect reflected photons. 
This approach stabilizes the excitation path and improves the efficiency of the atom-photon interface, allowing for deterministic interactions with single itinerant photons.
High-contrast reflection measurements are enabled by the high-cooperativity, critically-coupled atom-cavity system. 
Resonant light is sent via the fiber network [Fig.~\ref{fig:setup}(b)] and reflected off of the target device. 
We pick off a small fraction ($\sim 10\%$) of this signal and use it to monitor the wide-band reflection spectrum on a spectrometer (Horiba iHR-550) as well as calibrate the coupling efficiency to the nanocavity. 
The remaining reflection is then routed either directly to a single-photon counting module (SPCM, Excelitas SPCM-NIR), 
or into a time-delay interferometer for use in spin-photon experiments [Sec.~\ref{sec:spinphoton}]. 
Due to this high-efficiency fiber-coupled network, we observe overall collection efficiencies of $\sim 40\%$, limited by the quantum efficiency of our APDs.

\section{Optimal strain regimes for SiV spin-photon experiments}
\label{sec:strain}

    \begin{figure}
		\includegraphics[width=\linewidth]{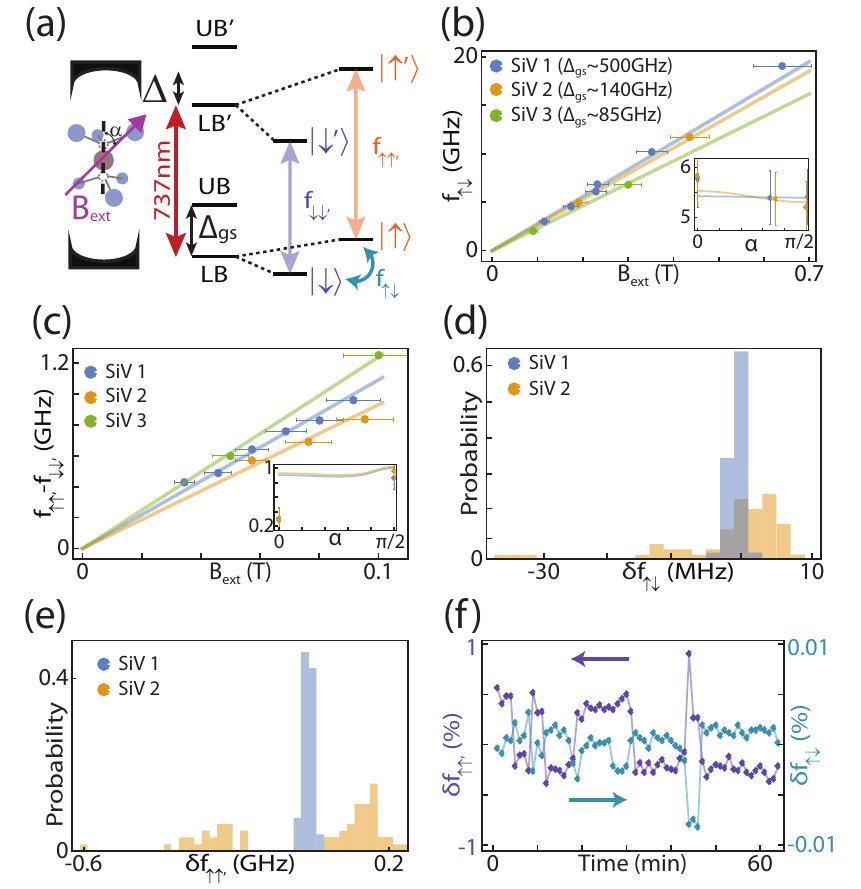}
	\caption{
			(a) SiV level diagram. Optical transitions $f_{\uparrow\uparrow'}, f_{\downarrow\downarrow'}\sim$\SI{737}{\nano\meter} are coupled to a nanophotonic cavity with mean detuning $\Delta$. Microwaves at frequency $f_{\uparrow\downarrow}$ drive rotations in the lower branch (LB). 
			(b) Qubit frequency $f_{\uparrow\downarrow}$ for differently strained emitters. Modeled splitting for ground state g-factors $\mathrm{g}_{\mathrm{gs}1} = 1.99, \mathrm{g}_{\mathrm{gs}2} = 1.89, \mathrm{g}_{\mathrm{gs}3} = 1.65$ (solid lines) based on independent measurements of $\Delta_{\mathrm{gs}}$. (inset) Angle dependence of $f_{\uparrow\downarrow}$ at fixed field $B_{\mathrm{ext}} = $ \SI{0.19}{\tesla}. Solid lines are predictions using the same model parameters.
			(c) Optical splitting $f_{\uparrow\uparrow'}-f_{\downarrow\downarrow'}$. Fits extract excited state g-factors $\mathrm{g}_{\mathrm{es}1} = 1.97, \mathrm{g}_{\mathrm{es}2} = 1.83, \mathrm{g}_{\mathrm{es}3} = 1.62$ (solid lines). (inset) Angle dependence of $f_{\uparrow\uparrow'}-f_{\downarrow\downarrow'}$ at fixed field $B_{\mathrm{ext}} = $ \SI{0.1}{\tesla}.
			(d) Histogram of MW transition frequency for two different emitters. 
			(e) Histogram of Optical transition frequency for two different emitters. 
			(f) Simultaneous measurement of $f_{\uparrow\downarrow}$ and $f_{\uparrow\uparrow'}$ reveals correlations between optical and microwave spectral diffusion for emitter 2.
			}
	    \label{fig:sivstrain}
    \end{figure}
		
Similar to other solid state emitters \cite{doherty2011negatively, Moison1994self}, the SiV is sensitive to local inhomogeneity in the host crystal. In the case of the SiV, which has $\mathrm{D}_{3d}$ symmetry, the dominant perturbation is crystal strain. In this section, we describe the effects of  strain on the SiV spin and optical properties, and how they can enable efficient microwave and optical control of SiV centers inside nanostructures.

\subsection{SiV Hamiltonian in the presence of strain and spin-orbit coupling}
The SiV electronic structure is comprised of spin-orbit eigenstates split by spin-orbit interactions. 
Optical transitions connect the ground state manifold ($LB$, $UB$) and excited state manifold ($LB'$, $UB'$) [Fig.~\ref{fig:sivstrain}(a)]. 
In a DR, phonon absorption $LB\rightarrow UB$ (and $LB'\rightarrow UB'$) is supressed, resulting in thermal polarization into $LB$. 

We consider the ground state SiV Hamiltonian with spin-orbit and strain interactions, in the combined orbital and spin basis $\{|e_y\uparrow\rangle, |e_y\downarrow\rangle, |e_x\uparrow\rangle, |e_x\downarrow\rangle\}$ \cite{hepp2014electronic, meesala2018strain}
\begin{align}
\mathcal{H}_{SiV} &= \mathcal{H}_{SO} + \mathcal{H}_{\textrm{strain}}\\
 &= 
\begin{pmatrix}
\alpha-\beta& 0& \gamma-i\lambda& 0\\
0& \alpha-\beta& 0& \gamma+i\lambda\\
\gamma+i\lambda& 0& \alpha+\beta& 0\\
0& \gamma-i\lambda& 0& \alpha+\beta
\end{pmatrix}
\label{eq:strainH}
\end{align}
 where $\alpha$ corresponds to axial strain, $\beta$ and $\gamma$ correspond to transverse strain, and $\lambda$ is the strength of spin-orbit interaction. 
Diagonalizing this reveals the orbital character of the lower branch: 
\begin{equation}
LB \propto 
\begin{cases}
|e_x\uparrow\rangle - \frac{1 + \sqrt{1 + (\gamma/\beta)^2 + (\lambda/\beta)^2}}{\gamma/\beta -i\lambda/\beta} |e_y\uparrow\rangle \\
|e_x\downarrow\rangle - \frac{1 - \sqrt{1 + (\gamma/\beta)^2 + (\lambda/\beta)^2}}{\gamma/\beta -i\lambda/\beta} |e_y\downarrow\rangle 
\end{cases}
\end{equation}
We investigate these electronic levels in the context of the SiV as a spin-photon interface.

\subsection{Effects of strain on the SiV qubit states}
In the limit of zero crystal strain, the orbital factors simplify to the canonical form \cite{hepp2014electronic} 
\begin{equation}
LB = 
\begin{cases}
|e_+\downarrow\rangle \\
|e_-\uparrow\rangle
\end{cases}
\end{equation}
In this regime, the spin-qubit has orthogonal electronic orbital and spin components. 
As result, one would need to simultaneously drive an orbital and spin flip to manipulate the qubit, which is forbidden for direct microwave driving alone. 
Thus, in the low strain regime, two-photon optical transitions between the qubit states in a misaligned external field, already demonstrated at millikelvin temperatures in \cite{becker2018all}, are likely necessary to realize a SiV spin qubit.

In the high strain limit ($\sqrt{\beta^2+\gamma^2}\gg\lambda$), these orbitals become 
\begin{equation}
LB = 
\begin{cases}
(\cos(\theta/2)|e_{x}\rangle-\sin(\theta/2)|e_{y}\rangle)\otimes|\downarrow\rangle\\
 (\cos(\theta/2)|e_{x}\rangle-\sin(\theta/2)|e_{y}\rangle)\otimes|\uparrow\rangle
\end{cases}
\end{equation}
where $\tan(\theta)= \frac{\beta}{\gamma}$. In this regime, the ground state orbital components are identical, and the qubit states can be described by the electronic spin degree of freedom only.
As such, the magnetic dipole transition between the qubit states is now allowed and can be efficiently driven with microwaves.

In addition to determining the efficiency of qubit transitions, the spin-orbit nature of the SiV qubit states also determines its susceptibility to external fields. 
%
%
In an externally applied magnetic field, $LB$ splits due to magnetic moments associated with both spin\textendash\ and orbital\textendash\ angular momenta. 
This splitting is parameterized by an effective g-tensor which, for a fixed angle between the external field and the SiV symmetry axis, can be simplified to an effective g-factor: $\mu\,\mathrm{g}\,B_{\mathrm{ext}}/h = f_{\uparrow\downarrow}$. 
In the limit of large strain, the orbital component of the two $LB$ wavefunctions converge, and $\mathrm{g}$ trends towards that of a free electron ($\mathrm{g}=2$). 
As a result, the qubit states behave akin to a free-electron in the high strain regime, and there is no dependence of $\mathrm{g}$ on external field angle or small perturbations in crystal strain. 

While it is difficult to probe $\beta$ or $\gamma$ directly, they relate to the energy difference between $UB$ and $LB$ via $\Delta_{\mathrm{gs}} = 2\sqrt{\beta_{\mathrm{gs}}^2+\gamma_{\mathrm{gs}}^2+\lambda_{\mathrm{gs}}^2}$ [Fig.~\ref{fig:sivstrain}(a)]. 
From this, we extract $\sqrt{\beta^2+\gamma^2}$, given the known value of $\lambda_{\mathrm{gs}}=$\SI{46}{\giga\hertz} \cite{hepp2014electronic, muller2014optical, rogers2014all}. 
Numerically diagonalizing the SiV Hamiltonian using the extracted values for $\beta$ and $\gamma$ closely matches the measured ground state splitting, both as a function of applied field magnitude and angle [Fig.~\ref{fig:sivstrain}(b)]. 

\subsection{Effects of strain on the SiV spin-photon interface}
Strain also plays a crucial role in determining the optical interface to the SiV spin qubit. 
The treatment shown above can be repeated for the excited states, with the caveat that the parameters $\beta, \gamma,$ and $\lambda$ are different in the excited state manifold as compared to the ground state manifold \cite{meesala2018strain}. 
These differences give rise to a different g-factor in the excited state ($\mathrm{g}_{\mathrm{es}}$). 
If the strain is much larger than both $\lambda_{\mathrm{gs}}=$\SI{46}{\giga\hertz} and $\lambda_{\mathrm{es}}=$\SI{255}{\giga\hertz}, then $\mathrm{g}_{\mathrm{gs}}\approx \mathrm{g}_{\mathrm{es}}\approx 2$. 
In this case, the two spin-cycling transition frequencies $f_{\uparrow\uparrow'}$ and $f_{\downarrow\downarrow'}$ are identical, and the only spin-selective optical transitions are the dipole-forbidden spin-flipping transitions $f_{\uparrow\downarrow'}$ and $f_{\downarrow\uparrow'}$.

Under more moderate strain, the difference $\delta \mathrm{g} = |\mathrm{g}_{\mathrm{es}}-\mathrm{g}_{\mathrm{gs}}|$ splits the degenerate optical transitions $f_{\uparrow\uparrow'}$ and $f_{\downarrow\downarrow'}$, making them spin-selective as well.
Due to differences in the anisotropic g-tensor in the ground and excited states, $\delta \mathrm{g}$ depends on the orientation of the magnetic field as well, and is minimized in the case of a $\langle 111 \rangle$-aligned field [Fig~\ref{fig:sivstrain}(c), inset]. 
In such an external field aligned with the SiV symmetry axis, optical transitions become highly spin-conserving \cite{sukachev2017silicon}, allowing many photons to scatter without altering the SiV spin state. 
This high cyclicity enables high-fidelity single-shot readout of the spin state \cite{nguyen2019quantum}, even without high collection efficiencies \cite{sukachev2017silicon}. 
This makes working with the spin-cycling transitions highly desirable, at the expense of a reduced ability to resolve spin-selective transitions for a given field magnitude.
The need to resolve individual transitions suggests an optimal strain regime where $\sqrt{\beta_{\mathrm{gs}}^2+\gamma_{\mathrm{gs}}^2}\gg \lambda_{\mathrm{gs}}$, where MW driving is efficient, while $\sqrt{\beta_{\mathrm{es}}^2+\gamma_{\mathrm{es}}^2}\lesssim \lambda_{\mathrm{es}}$, where one can independently address $f_{\uparrow\uparrow'}$ and $f_{\downarrow\downarrow'}$ [Fig.~\ref{fig:sivstrain}(c)]. 

\subsection{Effects of strain on SiV stability}

Despite the SiV's symmetry-protected optical transitions, spectral diffusion of the SiV has been observed in many experiments \cite{rogers2014multiple, evans2016narrow} (but still much smaller compared to emitters without inversion symmetry, for example, nitrogen-vacancy centers \cite{faraon2012coupling, riedel2017deterministic}). 
While the exact nature of this diffusion has not been studied in depth, it is often attributed to the second-order Stark effect or strain fluctuations, both of which affect the energies of SiV orbital wavefunctions. 
In this paper, we also observe significant fluctuations of the spin qubit frequency.

As can be seen in reference \cite{meesala2018strain}, for an appropriately low static strain value, fluctuating strain can give rise to fluctuations in the g-tensor of the ground state, causing spectral diffusion of the qubit frequency $f_{\uparrow\downarrow}$ [Fig.~\ref{fig:sivstrain}(d)]. 
Since $\mathrm{g}_{\mathrm{gs}}$ asymptotically approaches 2 as the static strain increases~\cite{meesala2018strain}, the qubit susceptibility to this fluctuating strain is reduced in the case of highly strained SiV centers, resulting in a more stable qubit. 

While spectral diffusion of the optical transition should not saturate in the same way as diffusion of the microwave transition, we observe qualitatively different spectral diffusion properties for different emitters [Fig.~\ref{fig:sivstrain}(e) and Fig.~\ref{fig:diffusion}]. 
SiV 1 ($\Delta_{\mathrm{gs}} =$ \SI{500}{\giga\hertz}) displays slow drift of the optical line which is stable to $<$\SI{100}{\mega\hertz} over many minutes [Appendix.~\ref{apx:diffusion}]. 
We do not observe significant fluctuations ($ > $ \SI{500}{kHz}) of the microwave transition for this SiV.
On the other hand, SiV 2 ($\Delta_{\mathrm{gs}} =$ \SI{140}{\giga\hertz}) drifts over a wider range, and also exhibits abrupt jumps between several discrete frequencies [Appendix.~\ref{apx:diffusion}]. 

We simultaneously record the optical transition and qubit frequency for SiV 2 and observe correlations between the two frequencies [Fig.~\ref{fig:sivstrain}(f)], indicating that they could arise from the same environmental perturbation.
In Appendix \ref{apx:strain}, we calculate the qubit and optical transition frequencies using the strain Hamiltonian (eq.~\ref{eq:strainH}) and find that both correlations and absolute amplitudes of spectral diffusion can simultaneously be explained by strain fluctuations on the order of $1\%$ ($\sim 10^{-7}$ strain) [Appendix~\ref{apx:strain}].

In this work we rely on static strain, likely resulting from damage induced by ion implantation and nanofabrication, and select for spectrally stable SiVs with appropriate strain profiles.
This is characterized by first measuring $\Delta_{gs}$ in zero magnetic field at \SI{4}{\kelvin} by exciting the optical transition $LB \rightarrow LB'$ and measuring emission from the $LB' \rightarrow UB$ on a spectrometer. 
We use this to screen for SiVs with $\Delta_{gs}>$\SI{100}{\giga\hertz} to ensure efficient MW driving of the spin qubit. 
We further apply a static external magnetic field and measure spectral stability properties as well as $f_{\uparrow\uparrow'} - f_{\downarrow\downarrow'}$ to guarantee a good spin-photon interface. We measured $\sim 10$ candidate emitters, and found $4$ which satisfy all of the necessary criteria for spin-photon experiments. 

\section{Regimes of cavity-QED for SiV spin-photon interfaces}
\label{sec:cavity}
    \begin{figure}
		\includegraphics[width=\linewidth]{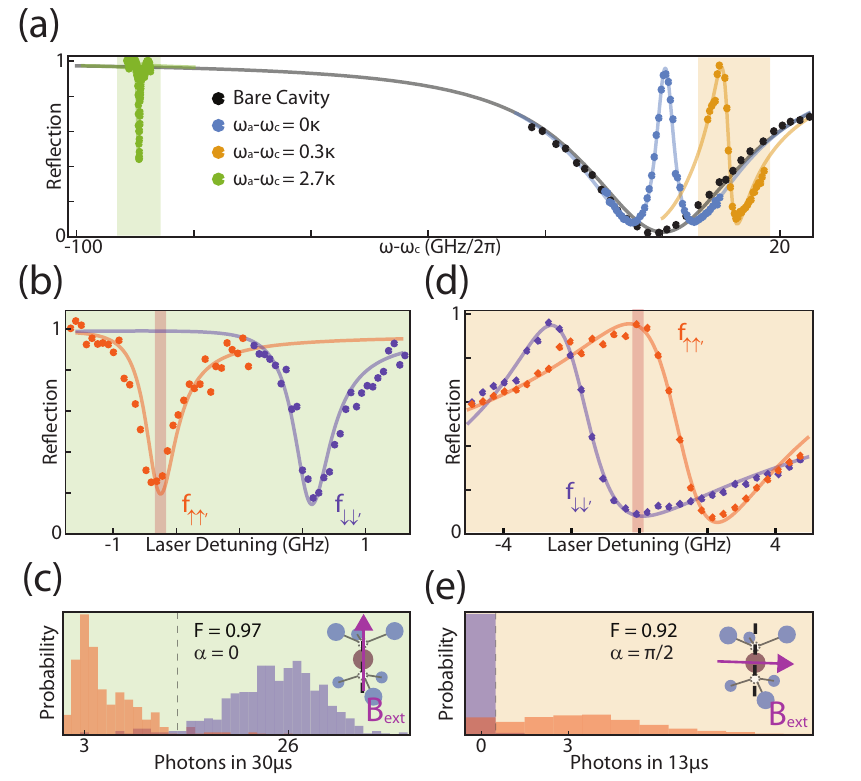}
	\caption{
			(a) SiV-cavity reflection spectrum at several detunings. The bare cavity spectrum (black) is modulated by the presence of the SiV. When the atom cavity detuing is small (Blue, orange), high-contrast, broad features are the result of Purcell enhanced SiV transitions. Far from the cavity resonance (green), interaction results in narrow SiV-assisted transmission channels.
			(b) Spin-dependent reflection for large SiV-cavity detuning $\Delta \approx -3\kappa$, $B_{\mathrm{ext}} =$ \SI{0.35}{\tesla}. In this regime, SiV spin states can be individually addressed.
			(c) Probing either transmission dip results in high-fidelity single-shot readout in an aligned field ($F=0.97$, threshold on detecting 13 photons).
			(d) Spin-dependent reflection near resonance $\Delta \approx 0.5\kappa$, $B_{\mathrm{ext}} =$ \SI{0.19}{\tesla}. Dispersive lineshapes allow for distinguishable reflection spectra from both SiV spin states.
			(e) A probe at the frequency of maximum contrast ($f_Q$) can determine the spin state in a single shot in a misaligned field($F=0.92$, threshold on detecting $>1$ photon).
						}
	    \label{fig:sivcavity}
    \end{figure}
Efficient spin-photon interactions are enabled by incorporating SiV centers into nanophotonic cavities. In this section, we describe SiV-cavity measurements in several regimes of cavity QED, and comment on their viability for spin-photon experiments.

\subsection{Spectroscopy of cavity-coupled SiVs}

We measure the spectrum of the atom-cavity system at different atom-cavity detunings in order to characterize the device and extract key cavity QED parameters [Fig.~\ref{fig:sivcavity}(a)]. 
The reflection spectrum of a two-level system coupled to a cavity is modeled by solving the frequency response of the standard Jaynes-Cummings Hamiltonian using input-output formalism for a cavity near critical coupling \cite{reiserer2015cavity}: 
\begin{equation}
\mathcal{R}(\omega) = \left|1-\frac{2\kappa_l}{i(\omega-\omega_c) + \kappa_{\mathrm{tot}} + g^2/(i(\omega-\omega_a) + \gamma)} \right|^2,
	    \label{eq:sivcavity}
\end{equation}
where $\kappa_l$ is the decay rate from the incoupling mirror, $\kappa_{\mathrm{tot}}$ is the cavity linewidth, $\omega_c (\omega_a)$ is the cavity (atom) resonance frequency, $g$ is the single-photon Rabi frequency, and $\gamma$ is the bare atomic linewidth. 
Interactions between the SiV optical transition and the nanophotonic cavity result in two main effects. First, the SiV center can modulate the reflection spectrum of the bare cavity, as seen in the colored curves of figure \ref{fig:sivcavity}(a). Second, the coupling to the cavity can broaden the linewidth of the SiV based on the Purcell effect: $$\Gamma \approx \gamma + 4g^2/\kappa \frac{1}{1+4(\omega_c-\omega_a)^2/\kappa^2}$$ 
When the cavity is far detuned from the atomic transition $|\omega_c-\omega_a|\equiv \Delta > \kappa$ [Fig.~\ref{fig:sivcavity}(a), green], Purcell enhancement is negligible and the cavity and atomic linewidths ${\kappa, \gamma} = 2\pi\times \{33, 0.1\}$ \si{\giga\hertz} are estimated. 
When the cavity is on resonance with the atom ($\Delta = 0$), we fit (\ref{eq:sivcavity}) using previously estimated values of $\kappa$ and $\gamma$ to extract $g = 2\pi\times 5.6$ \si{\giga\hertz}. 
Together, these measurements allow us to determine the atom-cavity cooperativity $C = 4g^2/\kappa\gamma = 38$. 
Importantly, interactions between the SiV and single photons becomes deterministic when $C>1$.

As mentioned in section [Sec.~\ref{sec:strain}], we would like to make use of spectrally resolved spin conserving optical transitions ($f_{\uparrow\uparrow'}, f_{\downarrow\downarrow'}$) to build a spin-photon interface using the SiV. 
Here, we make this criteria more explicit: $f_{\uparrow\uparrow'}$ and $f_{\downarrow\downarrow'}$ can be resolved when $|f_{\uparrow\uparrow'} - f_{\downarrow\downarrow'}| \gtrsim \Gamma$.

\subsection{Cavity QED in the detuned regime}

In the detuned regime ($\Delta > \kappa$), $\Gamma \approx \gamma$, and narrow atom-like transitions are easily resolved under most magnetic field configurations, including when the field is aligned with the SiV symmetry axis [Fig.~\ref{fig:sivcavity}(b)]. 
In this case [sec.~\ref{sec:strain}] \cite{sukachev2017silicon}, optical transitions are highly spin-conserving, and many photons can be collected allowing for high-fidelity single-shot readout of the SiV spin state ($F=0.97$) [Fig.~\ref{fig:sivcavity}(c)]. 
Rapid, high-fidelity, non-destructive single-shot readout can enable projective-readout based initialization: after a single measurement of the SiV spin state, the probability of a measurement-induced spin flip is low, effectively initializing the spin into a known state.

While this regime is useful for characterizing the system, the maximum fidelity of spin-photon entanglement based on reflection amplitude is limited. 
As seen in figure \ref{fig:sivcavity}(b), the contrast in the reflection signal between an SiV in $\ket{\uparrow}$ (orange) vs. $\ket{\downarrow}$ (purple) is only $\sim80\%$, implying that in $20\%$ of cases, a photon is reflected from the cavity independent of the spin state of the SiV, resulting in errors. 
We note that the residual $20\%$ of reflection can be compensated by embedding the cavity inside an interferometer at the expense of additional technical stabilization challenges, discussed below.

\subsection{Cavity QED near resonance}

Tuning the cavity onto the atomic resonance ($\Delta \approx 0$) dramatically improves the reflection contrast [Fig.~\ref{fig:sivcavity}(a) (blue curve)]. 
Here, we observe nearly full contrast of the reflection spectrum due to the presence of the SiV. 
Unfortunately, this is associated with a broadened atomic linewidth ($\Gamma = \gamma (1+C) \sim$\SI{4}{\giga\hertz}). 
While it is, in principle, still possible to split the atomic lines by going to higher magnetic fields, there are several technical considerations which make this impractical. 
Large magnetic fields ($|B_{\mathrm{ext}}|>$\SI{0.5}{\tesla}) correspond to large qubit frequencies ($f_{\uparrow\downarrow}$), which can induce spontaneous qubit decay due to phonon emission ($\ket{\uparrow}\rightarrow\ket{\downarrow}$), as well as increased local heating of the device from microwave dissipation, both of which reduce the SiV spin coherence time rendering it ineffective as a quantum memory. 

At intermediate detunings ($0<\Delta<\kappa$), the SiV resonance is located on the cavity slope and results in high-contrast, spin-dependent Fano lineshapes which exhibit sharp features smaller than $\Gamma$ [Fig.~\ref{fig:sivcavity}(a), orange curve]. 
By working at an optimal $B_{\mathrm{ext}}$ where the peak of one spin transition is overlapped by the valley of the other, the best features of the resonant and far-detuned regimes are recovered [Fig.~\ref{fig:sivcavity}(e)]. 
Probing the system at the point of maximum contrast ($f_Q \approx (|f_{\uparrow\uparrow'} - f_{\downarrow\downarrow'}|)/2$, contrast $>90\%$) enables single-shot readout of the SiV spin state for an arbitrary field orientation, even when transitions are not cycling [Fig.~\ref{fig:sivcavity}(f)]. 

This demonstrates an optical regime of cavity QED where we simultaneously achieve high-contrast readout while maintaining spin-dependent transitions. 
In this regime, we still expect residual reflections of about $10\%$, which end up limiting spin-photon entanglement fidelity. 
This infidelity arises because the cavity is not perfectly critically coupled ($\kappa_l \neq \kappa_{tot}/2$), and can in principle be solved by engineering devices that are more critically coupled.
Alternatively, this problem can be addressed for any cavity by interfering the signal with a coherent reference to cancel unwanted reflections. 
In this case, one would have to embed the cavity in one arm of a stabilized interferometer. 
This is quite challenging, as it involves stabilizing $\sim $ \SI{10}{\meter} long interferometer arms, part of which lie inside the DR (and experience strong vibrations from the pulse-tube cryocooler).
A fundamental issue with critically coupled cavities is that not all of the incident light is reflected from the device. If the spin is not initialized in the highly-reflecting state, photons are transmitted and not recaptured into the fiber network. 
Switching to overcoupled (single-sided) cavities, where all photons are reflected with a spin-dependent phase, could improve both the fidelity and efficiency of spin-photon entanglement. 
Once again, however, measurement of this phase would require embedding the cavity inside of a stabilized interferometer. 
As such, the un-compensated reflection amplitude based scheme employed here is the most technically simple approach to engineering spin-photon interactions. 
\section{Microwave spin control}
\label{sec:mwcontrol}
    \begin{figure}
		\includegraphics[width=\linewidth]{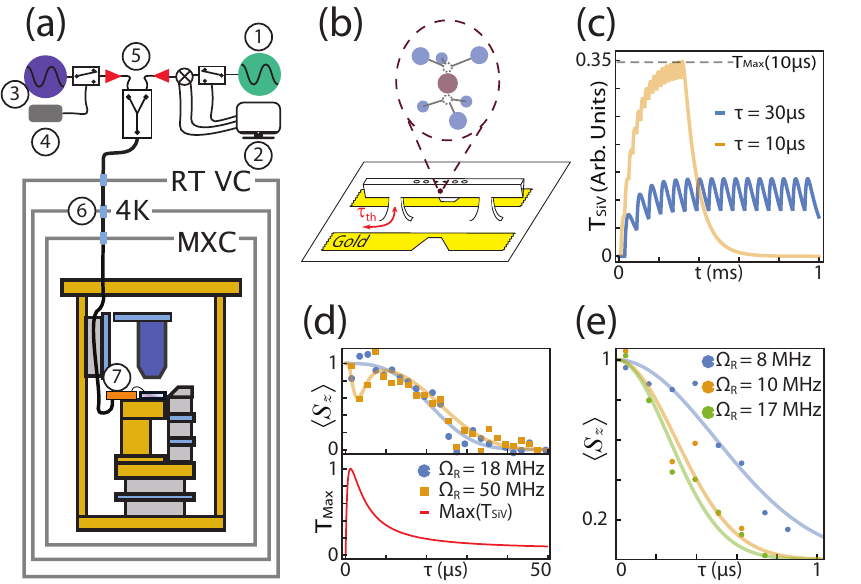}
	\caption{
			(a) Experimental schematic for microwave control. The amplitude and phase of a CW microwave source \protect\circled{\scriptsize{1}} are modulated via a microwave switch and IQ mixer controlled externally by an AWG \protect\circled{\scriptsize{2}}. A CW radio frequency source \protect\circled{\scriptsize{3}} is controlled using a digital delay generator \protect\circled{\scriptsize{4}}. Both signals are amplified by $30$dB amplifiers \protect\circled{\scriptsize{5}} before entering the DR. $0$dB cryo-attenuators \protect\circled{\scriptsize{6}} thermalize coax cables at each DR stage,  ultimately mounted to a PCB \protect\circled{\scriptsize{7}} on the sample stage and delivered to the devices.
			(b) Schematic depicting microwave-induced heating of devices.  
			(c) Modeled temperature at the SiV from a dynamical decoupling sequence. At long $\tau$, device cools down between each decoupling pulse, resulting in low temperatures. At short $\tau$, devices are insufficiently cooled, resulting in a higer max temperature ($T_{\mathrm{max}}$). 
			(d) Effects of microwave heating on SiV coherence time. (Top panel) At high Rabi frequencies, SiV coherence is temporarily reduced for small $\tau$. (Bottom panel) The local temperature ($T_{\mathrm{max}}$) at the SiV calculated by taking the maximum value of the plots in figure (c). 
			(e) Hahn-echo for even lower Rabi frequencies, showing coherence times that scale with microwave power
			}
	    \label{fig:mwheating}
    \end{figure}
While the optical interface described in previous sections enables high-fidelity initialization and readout of the SiV spin qubit, direct microwave driving is the most straightforward path towards coherent single-qubit rotations. 
Typically, microwave manipulation of electron spins requires application of significant microwave power. This presents a challenge, as SiV spins must be kept at local temperatures below \SI{500}{mK} in order to avoid heating-related dephasing. 
In this section, we implement coherent microwave control of SiV centers inside nanostructures at temperatures below \SI{500}{mK}. 

\subsection{Generating microwave single-qubit gates}
The SiV spin is coherently controlled using amplitude and phase controlled microwave pulses generated by a Hittite signal generator (HMC-T2220). 
A target pulse sequence is loaded onto an arbitrary waveform generator (Tektronix AWG 7122B), which uses a digital channel to control a fast, high-extinction MW-switch (Custom Microwave Components, CMCS0947A-C2), and the analog channels adjust the amplitude and phase via an IQ-mixer (Marki, MMIQ-0416LSM). 
The resulting pulse train is subsequently amplified (Minicircuits, ZVE-3W-183+) to roughly \SI{3}{\watt} of power, and sent via a coaxial cable into the dilution refrigerator. 
At each cryogenic flange, a \SI{0}{dB} attenuator is used to thermalize the inner and outer conductors of the coaxial line while minimizing microwave dissipation. 
The signal is then launched into a coplanar waveguide on a custom-built circuit board (Rogers4003C, Bay Area Circuits) so it can be wire-bonded directly to the diamond chip [Sec.~\ref{sec:fab}, Fig.~\ref{fig:mwheating}(c)]. 
The qubit frequency ($f_{\downarrow\uparrow}$) is measured by its optically detected magnetic resonance spectrum (ODMR) identically to the method described in \cite{sukachev2017silicon}. 
We observe ODMR from \SIrange{2}{20}{\giga\hertz} (corresponding to fields from \SIrange{0.1}{0.7}{\tesla}), implying that microwave control of SiV centers in this configuration is possible at a wide variety of external field magnitudes. 
This allows the freedom of tuning the field to optimize other constraints, such as for resolving spin transitions [Sec.~\ref{sec:cavity}] and identifying ancillary nuclear spins [Sec.~\ref{sec:Ncontrol}].

Once the qubit frequency has been determined for a given field, single-qubit gates are tuned up by measuring Rabi oscillations. 
The frequency of these oscillations scales with the applied microwave power $\Omega_R \sim \sqrt{P}$ and determines the single-qubit gate times. 
We can perform $\pi$-pulses ($R_{\phi}^{\pi}$) in under \SI{12}{\nano\second}, corresponding to a Rabi frequency exceeding \SI{80}{\mega\hertz} \cite{nguyen2019quantum}. 
This coherent control is used to implement pulse-error correcting dynamical decoupling sequences, either CPMG-N sequences of the form $R_x^{\pi/2} - \left(\tau - R_y^{\pi} - \tau\right)^N - R_x^{\pi/2} = x-(Y)^N-x$\cite{meiboom1958cpmg} or XY8-N sequences of the form $x-(XYXYYXYX)^N-x$ \cite{gullion1990xy8}. 
Sweeping the inter-pulse delay $\tau$ measures the coherence time $T_2$ of the SiV. 

\subsection{Effects of microwave heating on coherence}
As mentioned in sections \ref{sec:setup} and \ref{sec:strain}, thermally induced $T_1$ relaxation can dramatically reduce SiV coherence times. 
To explain this phenomenon, we model the nanobeam as a $1D$ beam weakly coupled at two anchor points to a uniform thermal bath [Fig.~\ref{fig:mwheating}(b)]. 
Initially, the beam is at the steady-state base temperature of the DR. 
A MW pulse instantaneously heats the bath, and the beam rethermalizes on a timescale $\tau_{\mathrm{th}}$ set by the thermal conduction of diamond and the beam geometry. 
Once the pulse ends, this heat is extracted from the beam on a similar timescale. 
By solving the time-dependent 1-D heat equation, we find that the change in temperature at the SiV caused by a single pulse (starting at time $t_0$) scales as $T_{\mathrm{SiV}} \propto (e^{-(t-t_0)/\tau_{\mathrm{th}}} - e^{-9(t-t_0)/\tau_{\mathrm{th}}})$. 
We take the sum over $N$ such pulses to model the effects of heating from a dynamical-decoupling sequence of size $N$.

At early times ($\tau<\tau_{\mathrm{th}}$), the SiV does not see the effects of heating by the MW line, and coherence is high. 
Similarly, at long times ($\tau\gg \tau_{\mathrm{th}}$) a small amount of heat is able to enter the nanostructure and slightly raise the local temperature, but this heat is dissipated before the next pulse arrives [Fig.~\ref{fig:mwheating}(c), blue curve]. 
At intermediate timescales however, a situation can arise where the nanobeam has not fully dissipated the heat from one MW pulse before the second one arrives [Fig.~\ref{fig:mwheating}(c), orange curve]. 
We plot the maximum temperature as seen by the SiV as a function of pulse spacing [Fig.~\ref{fig:mwheating}(d), lower panel], and observe a spike in local temperature for a specific inter-pulse spacing $\tau$, which depends on $\tau_{\mathrm{th}}$. 
Dynamical-decoupling sequences using high Rabi frequency pulses reveal a collapse in coherence at a similar time [Fig.~\ref{fig:mwheating}(d), upper panel]. 
This collapse disappears at lower Rabi frequencies, suggesting that it is associated with heating-related dephasing. 
We fit this collapse to a model where the coherence time $T_2$ depends on temperature \cite{jahnke2015electron}, and extract the rate of heating $\tau_{\mathrm{th}} =$ \SI{70}{\micro\second}.

Typically, faster $\pi$-pulses improve measured spin coherence by minimizing finite-pulse effects and detuning errors. 
Unfortunately, as seen above, faster pulses require higher MW powers which cause heating-related decoherence in our system. 
We measure Hahn-echo at lower MW powers [fig.~\ref{fig:mwheating}(e)], and find MW heating limits $T_2$ even at $\Omega_R \sim$\SI{10}{\mega\hertz}. 
For applications where long coherence is important, such as electron-nuclear gates [Sec.~\ref{sec:Ncontrol}], we operate at an optimal Rabi frequency $\Omega_R = 2\pi\times$\SI{10}{\mega\hertz} where nuclear gates are as fast as possible while maintaining coherence for the entire gate duration. 
For applications such as spin-photon entangling gates where fast gates are necessary [Sec.~\ref{sec:spinphoton}], we operate at higher Rabi frequencies $\Omega_R = 2\pi\times$\SI{80}{\mega\hertz} at the cost of reduced coherence times. 

Heating related effects could be mitigated by using superconducting microwave waveguides. 
This approach would also enable the fabrication of a single, long superconducting waveguide that could simultaneously address all devices on a single chip. 
However, it is still an open question whether or not superconducting waveguides with appropriate critical temperature, current, and field properties can be fabricated around diamond nanostructures. 

\section{Investigating the noise bath of SiVs in nanostructures}
\label{sec:mwdeer}

    \begin{figure}
		\includegraphics[width=\linewidth]{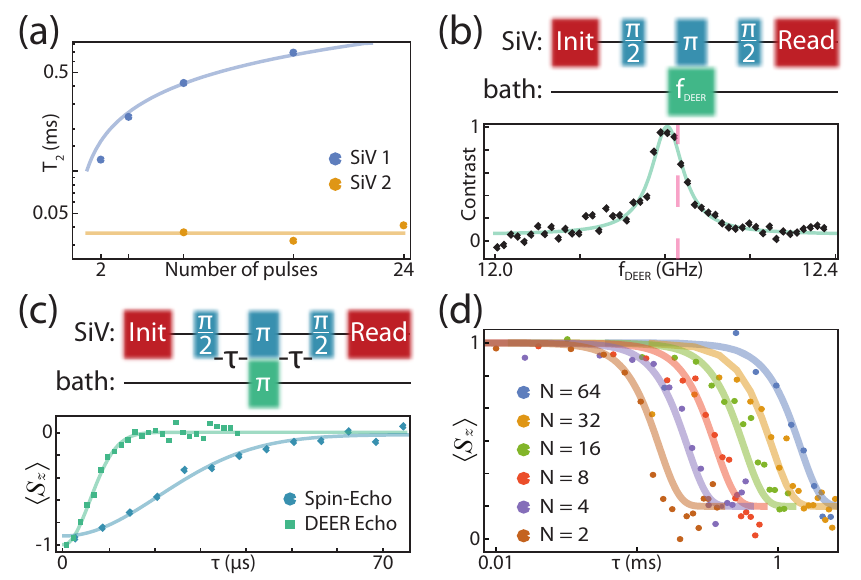}
	\caption{
			(a) T2 scaling for two different SiVs. SiV 2 exhibits no scaling with number of pulses ($T_{2,\mathrm{SiV 2}} =$\SI{30}{\micro\second}).
			(b) DEER ESR on SiV 2. Vertical red line is the expected frequency of a $\mathrm{g}=2$ spin based on our ability to determine the applied external field (Typically to within $10\%$).
			(c) DEER Echo on SiV 2. $T_{2,DEER} =$\SI{10}{\micro\second}.
			(d) Dynamical-decoupling on SiV 1. Data points are $T_2$ measurements used in part (a, blue curve), and solid lines are a noise model consisting of two Lorentzian noise baths.
			}
	    \label{fig:mwdeer}
    \end{figure}
At low temperatures, the coherence time of SiV centers drastically depends on the surrounding spin bath, which can differ from emitter to emitter. 
As an example, we note that the $T_2$ of two different SiV centers in different nanostructures scales differently with the number of applied decoupling pulses [Fig.~\ref{fig:mwdeer}(a)]. 
Surprisingly, the coherence time of SiV 2 does not scale with the number of applied pulses, while the coherence time of SiV 1 does scale as $T_2(N) \propto N^{2/3}$. 
Notably, both scalings are different as compared to what was previously measured in bulk diamond: $T_2(N) \propto N^{1}$ \cite{sukachev2017silicon}. 
In this section, we probe the spin bath of these two SiVs in nanostructures to investigate potential explanations for the above observations.

\subsection{Double electron-electron resonance spectroscopy of SiVs in nanostructures}

In order to investigate the poor coherence of SiV 2, we perform double electron-electron resonance (DEER) spectroscopy \cite{Milov1984electron} to probe the spin bath surrounding this SiV. 
We perform a Hahn-echo sequence on the SiV, and sweep the frequency of a second microwave pulse (taking the RF path in figure \ref{fig:mwheating}(a)), contemporaneous with the echoing SiV $\pi$-pulse [Fig.~\ref{fig:mwdeer}(b), upper panel]. 
If this second pulse is resonant with a spin bath coupled to the SiV, the bath can flip simultaneously with the SiV, leading to increased sensitivity to noise from the bath [Fig.~\ref{fig:mwdeer}(b), lower panel].
We observe a significant reduction of coherence at a frequency consistent with that of a free-electron spin bath ($\mathrm{g}_{\mathrm{bath}} = 2$) (resonance expected at $12(1)$ \si{\giga\hertz}). 

Next, we repeat a standard Hahn-echo sequence where a $\pi$-pulse resonant with this bath is applied simultaneously with the SiV echo pulse (DEER echo). 
The coherence time measured in DEER echo is significantly shorter than for standard spin-echo, indicating that coupling to this spin bath is a significant source of decoherence for this SiV. 
One possible explanation for the particularly severe bath surrounding this SiV is a thin layer of alumina ($\mathrm{Al}_2\mathrm{O}_3$) deposited via atomic layer deposition on this device in order to tune cavities closer to the SiV transition frequency. 
The amorphous oxide layer\textendash or its interface with the diamond crystal\textendash can be host to a large number of charge traps, all located within $\sim$\SI{50}{\nano\meter} of this SiV. 
Unfortunately, we could not measure this device without alumina layer due to our inability to gas-tune the nanophotonic cavity close enough to the SiV resonance [Sec.~\ref{sec:setup}].

These observations are further corroborated by DEER measurements in SiV 1, where the alumina layer was not used (only $\mathrm{N_2}$ was used to tune this cavity). 
In this device, we observe longer coherence times which scale $T_2(N) \propto N^{2/3}$, as well as no significant signatures from $\mathrm{g}_{\mathrm{bath}} = 2$ spins using DEER spectroscopy. 
We fit this scaling to a model consisting of two weakly-coupled spin baths [Fig.~\ref{fig:mwdeer}(d), Appendix.~\ref{apx:bath}], and extract bath parameters $b_1 = $\SI{5}{\kilo\hertz}, $\tau_1 = $\SI{1}{\micro\second}, $b_2 = $\SI{180}{\kilo\hertz}, $\tau_2 = $\SI{1}{\milli\second}, where $b$ corresponds to the strength of the noise bath, and $\tau$ corresponds to the correlation time of the noise \cite{deLange2010Universal, Myers2014Probing}. 

While the source of this noise is an area of future study, we find that the $b_2$ term (likely due to bulk impurities) is the dominant contribution towards decoherence in the system [Appendix.~\ref{apx:bath}]. 
Removing this term from the model results in coherence times up to a factor of 1000 times larger than measured values. 
Higher-temperature \cite{evans2016narrow} or {\it in situ} \cite{Kalish1995graphitization} annealing could potentially mitigate this source of decoherence by eliminating paramagnetic defects such as vacancy clusters. 
Additionally, by accompanying Si implantation with electron irradiation \cite{Acosta2009diamonds}, SiV centers could be created more efficiently, and with reduced lattice damage. 
Finally, working with isotopically purified diamond samples with very few \cnuc, a spin-1/2 isotope of carbon, could also result in a reduced spin bath \cite{sukachev2017silicon}, [Appendix.~\ref{apx:bath}]. 
\section{Spin-photon entanglement}
\label{sec:spinphoton}
    \begin{figure}
		\includegraphics[width=\linewidth]{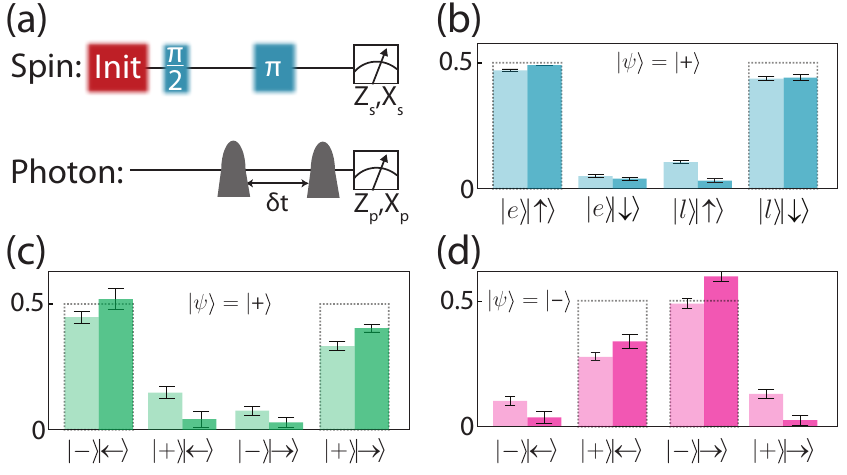}
	\caption{
			(a) Experimental sequence for generating and verifying spin-photon entanglement. A time-bin encoded qubit is reflected by the cavity, and both the SiV and the photonic qubits are measured in the Z and X bases.
			(b) Spin-photon correlations measured in the Z-Z basis. Light (dark) bars are before (after) correcting for known readout error associated with single-shot readout of the SiV.
			(c) Spin-photon correlations measured in the X-X basis. Bell-state preparation fidelity of $F \geq 0.89(3)$ and a concurrence $\mathcal{C} \geq 0.72(7)$.
			(d) Preparation of second spin-photon Bell state. Changing the phase of the incoming photonic qubit prepares a Bell-state with inverted statistics in the X basis. 
			}
	    \label{fig:spinphoton}
    \end{figure}
		
The previous sections characterize the SiV as an efficient spin-photon interface and a quantum memory with long-lived coherence. 
Here, we combine these two properties to demonstrate entanglement between a spin qubit and a photonic qubit. 
The mechanism for generating entanglement between photons and the SiV can be seen in figure \ref{fig:sivcavity}(b,d): Depending on the spin state of the SiV, photons at the probe frequency are either reflected from the cavity and detected, or are transmitted and lost. 

\subsection{Generating time-bin qubits}
We begin by explaining our choice of time-bin encoding for photonic qubits.
One straightforward possibility is to use the Fock state of the photon. However, it is extremely challenging to perform rotations on a Fock state, and photon loss results in an error in the computational basis.
Another, perhaps more obvious possibility is to use the polarization degree of freedom. While the SiV spin-photon interface is not polarization selective (both spin states couple to photons of the same polarization), one could consider polarization based spin-photon entangling schemes already demonstated in nanophotonic systems \cite{duan2004scalable, tiecke2014nanophotonic}. However, this requires embedding the nanostructure inside of a stabilized interferometer, which has a number of challenges [Sec.~\ref{sec:cavity}]. In addition, it requires careful fabrication of overcoupled, single-sided cavities (unlike the critically coupled diamond nanocavities used here [Sec.~\ref{sec:fab}]).
As such, we believe time-bin encoding is a natural choice given the critically-coupled SiV-cavity interface described here [Sec.~\ref{sec:cavity}].

These qubits are generated by passing a weak coherent laser though a cascaded AOM, amplitude-EOM, and phase-EOM. 
The time-bins are shaped by an AWG-generated pulse on the amplitude-EOM, and are chosen to be much narrower than the delay $\delta t$ between time bins. 
We can choose to prepare arbitrary initial photonic states by using the phase-EOM to imprint an optional phase shift to the second bin of the photonic qubit. 
Since we use a laser with Poissonian photon number statistics, we set the average photon number $\langle n_{ph}\rangle = 0.008 \ll 1$ using the AOM to avoid events where two photons are incident on the cavity.

Using this encoding, measurements in a rotated basis (X-basis) become straightforward. 
We send the time-bin qubit into an actively stabilized, unbalanced, fiber-based, Mach-Zender interferometer, where one arm passes through a delay line of time $\delta t$. 
With $25\%$ probability, $|e\rangle$ enters the long arm of the interferometer and $|l\rangle$ enters the short arm, and the two time bins interfere at the output. 
Depending on the relative phase between the two bins, this will be detected on only one of the two arms of the interferometer output [Fig.~\ref{fig:setup}(b)], corresponding to a measurement in the X basis of $|\pm\rangle$. 

\subsection{Spin-photon Bell states}
We prepare and verify the generation of maximally entangled Bell states between the SiV and a photonic qubit using the experimental sequence depicted in figure \ref{fig:spinphoton}(a). 
First, the SiV is initialized into a superposition state $\ket{\rightarrow} = 1/\sqrt2 (\ket{\uparrow}+ \ket{\downarrow})$. Then photons at frequency $f_Q$ [Sec.~\ref{sec:cavity}] are sent to the cavity, corresponding to an incoming photon state $|+\rangle = 1/\sqrt{2} (|e\rangle + |l\rangle)$, conditioned on the eventual detection of only one photon during the experiment run. 
Before any interactions, this state can be written as an equal superposition: $\Psi_{0} = |\rightarrow\rangle \otimes |+\rangle = 1/2 (|e\uparrow\rangle + |e\downarrow\rangle + |l\uparrow\rangle + |l\downarrow\rangle)$. 
The first time bin is only reflected from the cavity if the the SiV is in state $\ket{\uparrow}$, effectively carving out $|e\downarrow\rangle$ in reflection \cite{welte2017cavity}. 
A $\pi$-pulse on the SiV transforms the resulting state to $\Psi_1 = 1/\sqrt3 (|e\downarrow\rangle + |l\downarrow\rangle + |l\uparrow\rangle)$. 
Finally, reflection of the late time-bin off of the cavity carves out the state $|l\downarrow\rangle$, leaving a final entangled state $\Psi_2 = 1/\sqrt2 (|e\downarrow\rangle + |l\uparrow\rangle)$. 
To characterize the resulting state, we perform tomography on both qubits in the Z and X bases [Fig.~\ref{fig:spinphoton}(a)].

In order to enable high-bandwidth operation and reduce the requirements for laser and interferometric stabilization in generating and measuring time-bin qubits, it is generally beneficial to set $\delta t$ as small as possible. 
The minimum $\delta t$ is determined by two factors: First, each pulse must be broad enough in the time-domain (narrow enough in the frequency domain) so that it does not distort upon reflection off of the device. 
From figure \ref{fig:sivcavity}(d), the reflection specturm is roughly constant over a $\sim $ \SI{100}{MHz} range, implying that $\sim$ nanosecond pulses are sufficient. 
The second consideration is that a microwave $\pi$-pulse must be placed between the two pulses. In this experiment, we drive fast (\SI{12}{ns}) $\pi$-pulses. As such, we set $\delta t = $ \SI{30}{ns} and use \SI{5}{ns} optical pulses to satisfy these criteria.
\subsection{Spin-photon entanglement measurements}  
For Z-basis measurements, photons reflected from the cavity are sent directly to a SPCM and the time-of-arrival of the time-bin qubit is recorded. 
Afterwards, the SiV is read out in the Z-basis [Sec.~\ref{sec:cavity}]. 
Single-shot readout is calibrated via a separate measurement where the two spin-states are prepared via optical pumping and read out, and the fidelity of correctly determining the $|\uparrow\rangle$ ($\ket{\downarrow}$) state is $F_{\uparrow} = 0.85$ ($F_{\downarrow} = 0.84$), limited by the large 0 component of the geometric distribution which governs photon statistics for spin-flip systems [Sec.~\ref{sec:cavity}]. In other words, since we work in a misaligned field in this experiment, the probability of a spin flip is high, making it somewhat likely to measure 0 photons regardless of initial spin state.
Even before accounting for this known error [Appendix.~\ref{apx:fid}], we observe clear correlations between the photonic and spin qubits [Fig. \ref{fig:spinphoton}(b), light-shading]. 
Error bars for these correlation histograms (and the following fidelity calculations) are estimated by statistical bootstrapping, where the scattered photon histograms (post-selected on the detection of $|e\rangle$ or $|l\rangle$) are randomly sampled in many trials, and the variance of that ensemble is extracted. 

Measurements in the X-basis are performed similarly. The photon is measured through an interferometer as described above, where now the detector path information is recorded for the overlapping time-bin. 
After a $R_{y}^{\pi/2}$ pulse on the SiV, the scattered photon histograms again reveal significant correlations between the `+' and `-' detectors and the SiV spin state [fig.~\ref{fig:spinphoton}(c)]. 
By adding a $\pi$-phase between the early and late time bins, we can prepare an orthogonal Bell state. Measured correlations of this state are flipped in the X-basis [Fig.~\ref{fig:spinphoton}(d)]. 

Measurements of this Bell state in the Z- and X-bases are used to estimate a lower bound on the fidelity: $F = \langle \Psi^+|\rho|\Psi^+\rangle \geq 0.70(3)$ ($F \geq 0.89(3)$ after correcting for readout errors) [Appendix.~\ref{apx:fid}]. 
The resulting entangled state is quantified by its concrrence $\mathcal{C} \geq 0.42(6)$ ($\mathcal{C} \geq 0.79(7)$ after correcting for readout errors) [Appendix.~\ref{apx:fid}]. 
This high-fidelity entangled state between a photonic qubit and a quantum memory is a fundamental resource for quantum communication\cite{childress2005fault} and quantum computing schemes \cite{monroe2014large}, and can be used, for example, to demonstrate heralded storage of a photonic qubit into memory \cite{nguyen2019quantum}.

\section{Control of SiV-\cnuc\ register}
\label{sec:Ncontrol}

    \begin{figure}
		\includegraphics[width=\linewidth]{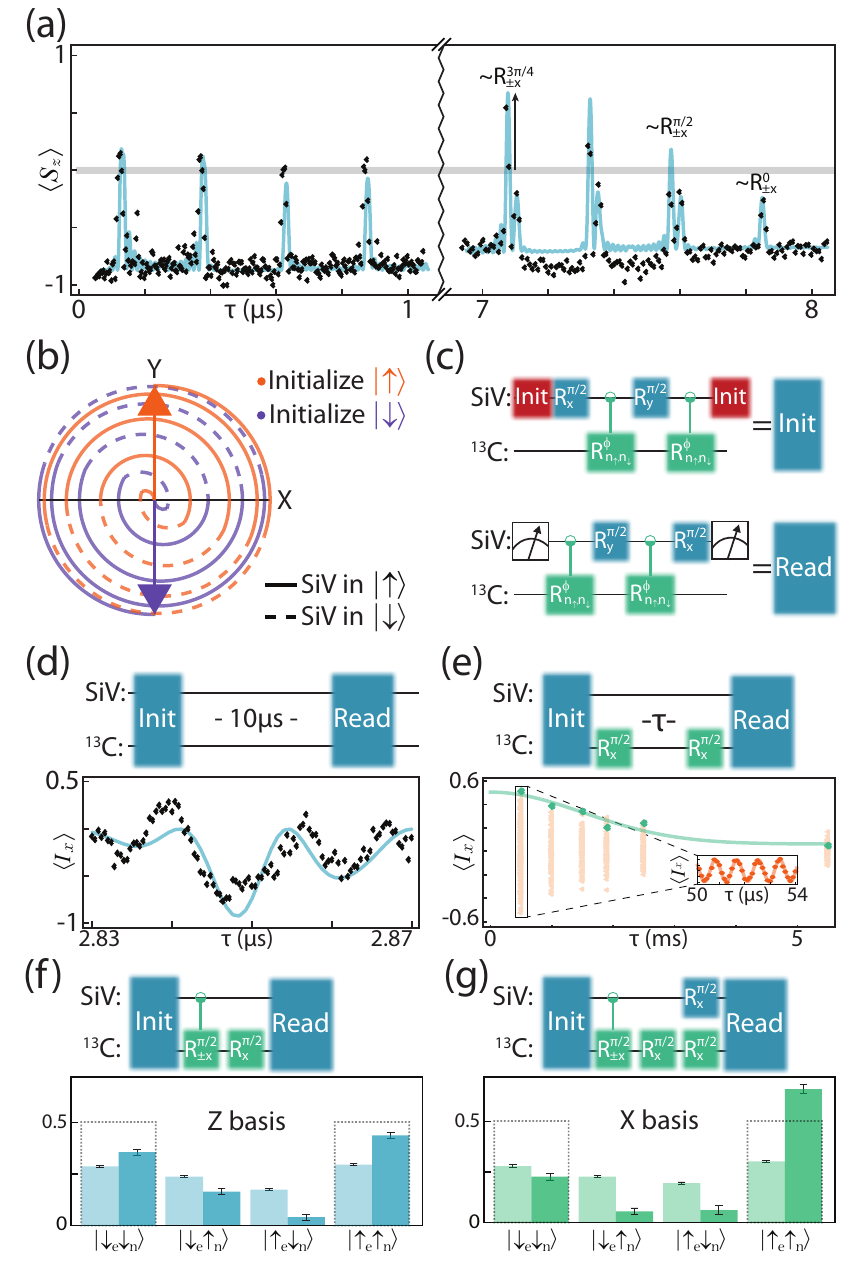}
	\caption{
			(a) XY8-2 spin echo sequence reveals coupling to nuclear spins. (Left panel) Collapses $\langle S_x\rangle = 0$ at short times indicate coupling to many nuclei. (Right panel) Collapses $\langle S_x\rangle \neq 0$ at long times indicate conditional gates on a single nuclear spin. 
			(b) Trajectory of $^{13}$C on the Bloch sphere during a maximally entangling gate. Orange (purple) lines correspond to the SiV initially prepared in state $|\uparrow \rangle$ $\left(|\uparrow \rangle \right)$; transitions from solid to dashed lines represent flips of the SiV electronic spin during the gate. 
			(c) Maximally entangling gates of the form $\mathcal{R}_{\vec{n_\uparrow},\vec{n_\downarrow}}^{\phi}$ are used to initialize and readout the two-qubit register. 
			(d) Tuning up an initialization gate. Inter-pulse spacing $\tau$ for Init and Read gates are swept to maximize polarization. Solid line is the modeled pulse sequence using the hyperfine parameters extracted from (a).
			(e) Nuclear Ramsey measurement. Driving the \cnuc\ using composite gates on the SiV reveals $T_2^* =$\SI{2.2}{\milli\second}. (Inset) Orange points are coherent oscillations of the Ramsey signal due to hyperfine coupling to the SiV.
			(f) Electron-nuclear correlations measured in the ZZ-basis. Light (dark) bars are before (after) correcting for known errors associated with reading out the SiV and \cnuc.
			(g) Electron-nuclear correlations measured in the XX-basis. We estimate a Bell state preparation fidelity of $F \geq 0.59(4)$ and a concurrence $\mathcal{C} \geq 0.22(9)$.
			}
	    \label{fig:ninit}
    \end{figure}
While demonstrations of a quantum node with a single qubit is useful for some protocol, nodes with several interacting qubits enable a wider range of applications, including quantum repeaters \cite{briegel1998repeater}. 
In this section, we introduce additional qubits based on \cnuc\ naturally occurring in diamond [sec.~\ref{sec:mwdeer}].

\subsection{Coupling between the SiV and several \cnuc}

For all of the emitters investigated in section \ref{sec:mwcontrol}, we observe collapses in the echo signal corresponding to entanglement with nearby nuclear spins [Fig.~\ref{fig:ninit}(a)]. 
As the diamond used in this work has $1\%$ \cnuc\ [Sec.~\ref{sec:mwdeer}], we typically observe several such nuclei, with all of their resonances overlapping due to their second-order sensitivity to hyperfine coupling parameters \cite{nguyen2019quantum}. 
Consequently, during a spin echo sequence the SiV entangles with many nuclei, quickly losing coherence and resulting in a collapse to $\langle S_z\rangle = 0$ [Fig.~\ref{fig:ninit}(a), left side]. 
If single \cnuc\ can be addressed however, this entanglement results in coherent population transfer and echo collapses which can, in some cases, completely flip the SiV spin state ($\langle S_z\rangle = \pm1$). 
This entanglement forms the basis for quantum gates [Fig.~\ref{fig:ninit}(a), right side]. 
These gates can be tuned by changing the alignment of $B_{\mathrm{ext}}$ with respect to the hyperfine coupling tensor, or by using different timings. 
Unfortunately, as a result of the complicated nuclear bath for this device, a majority of field orientations and amplitudes only show collapses to $\langle S_z\rangle = 0$. 
The highest fidelity nuclear gates demonstrated here are based on echo resonances with the largest contrast which, crucially, were not commensurate with an aligned field. Thus, in this device, single \cnuc\ could only be isolated at the cost of lower SSR fidelity [Sec.~\ref{sec:cavity}].

\subsection{Initializing the nuclear spin}

Once a single nuclear spin is identified, resonances in spin-echo form the building block for quantum gates. 
For example, a complete flip of the SiV is the result of the nuclear spin rotating by $\pi$ conditionally around the axes $\pm$X ($\mathcal{R}_{\pm x,\mathrm{SiV-C}}^{\pi}$), depending on the state of the SiV. 
We can vary the rotation angle of this pulse by choosing different spacings $\tau$ between pulses [Fig.~\ref{fig:ninit}(a)], or by using different numbers of $\pi$-pulses. We find a maximally entangling gate ($\mathcal{R}_{\pm x,\mathrm{SiV-C}}^{\pi/2}$) by applying $N = 8$ $\pi$-pulses separated by $2\tau = 2\times 2.859\,\mu$s.
This can be visualized on the Bloch sphere in figure \ref{fig:ninit}(b), where the state of the SiV (orange or purple) induces different rotations of the \cnuc.

A similarly constructed entangling gate ($\mathcal{R}_{\vec{n_\uparrow},\vec{n_\downarrow}}^{\phi}$, discussed in Appendix~\ref{apx:init}) is used to coherently map population from the SiV onto the nuclear spin or map population from the nuclear spin onto the SiV [Fig.~\ref{fig:ninit}(c)]. 
The fidelity of these gates is estimated by polarizing the SiV, mapping the population onto the \cnuc, and waiting for $T\gg T_2^*$ (allowing coherence to decay) before mapping the population back and reading out [Fig.~\ref{fig:ninit}(d)]. 
We find that we can recover $80\%$ of the population in this way, giving us an estimated initialization and readout fidelity of $F=0.9$. 
Based on the contrast of resonances in spin-echo (also $0.9$), this is likely limited by entanglement with other nearby \cnuc\ for this emitter, as well as slightly sub-optimal choices for $\tau$ and $N$. 
Coupling to other \cnuc\ results in population leaking out of our two-qubit register, and can be improved by increasing sensitivity to single \cnuc, or by looking for a different emitter with a different \cnuc\ distribution. 
The misaligned external field further results in slight misalignment of the nuclear rotation axis and angle of rotation, and can be improved by employing adapted control sequences to correct for these errors \cite{casanova2015robust, schwartz2018robust}. 

\subsection{Microwave control of nuclear spins}
As demonstrated above, control of the \cnuc\ via composite pulse sequences on the SiV is also possible. 
A maximally entangling gate has already been demonstrated and used to initialize the \cnuc, so in order to build a universal set of gates, all we require are unconditional single-qubit rotations. 
This is done following reference \cite{taminiau2014universal}, where unconditional nuclear rotations occur in spin-echo sequences when the inter-pulse spacing $\tau$ is halfway between two collapses. 
For the following gates, we use an unconditional $\pi/2$-pulse composed of 8 $\pi$-pulses separated by $\tau = $ \SI{0.731}{\micro\second}. 

We use this gate to probe the coherence time $T_2^*$ of the \cnuc. After mapping population onto the nuclear spin, the SiV is re-initialized, and then used to perform unconditional $\pi/2$-rotations on the \cnuc\ [Fig.~\ref{fig:ninit}(d)]. 
Oscillations in the signal demonstrate Larmor precession of the nucleus at a frequency determined by a combination of the external field as well as \cnuc-specific hyperfine interactions \cite{nguyen2019quantum}, which are seen as the orange data points in figure \ref{fig:ninit}(d). 
The green envelope is calculated by fitting the oscillations and extracting their amplitude. The decay of this envelope $T_2^* =$\SI{2.2}{\milli\second} shows that the \cnuc\ has an exceptional quantum memory, even in the absence of any dynamical decoupling. 

We characterize the fidelity of our conditional and unconditional nuclear gates by generating and reading out Bell states between the SiV and \cnuc\ [Appendix.~\ref{apx:fid}]. 
First, we initialize the 2-qubit register into one of the 4 eigenstates: $\{|\uparrow_e\uparrow_N\rangle, |\uparrow_e\downarrow_N\rangle, |\downarrow_e\uparrow_N\rangle, |\downarrow_e\downarrow_N\rangle\}$, then perform a $\pi/2$-pulse on the electron to prepare a superposition state. 
Afterward, a CNOT gate, comprised of an unconditional $\pi/2$ pulse followed by a maximally entangling gate, prepares one of the Bell states $|\Psi_{\pm}\rangle, |\Phi_{\pm}\rangle$ depending on the initial state [Fig. \ref{fig:ninit} (e,f)]. 
Following the analysis outlined in appendix \ref{apx:fid}, we report an error corrected fidelity of $F \geq 0.59(4)$ and $\mathcal{C} \geq 0.22(9)$, primarily limited by our inability to initialize the \cnuc\ \cite{nguyen2019quantum}.

\subsection{Radio-frequency driving of nuclear spins}
    \begin{figure}
		\includegraphics[width=\linewidth]{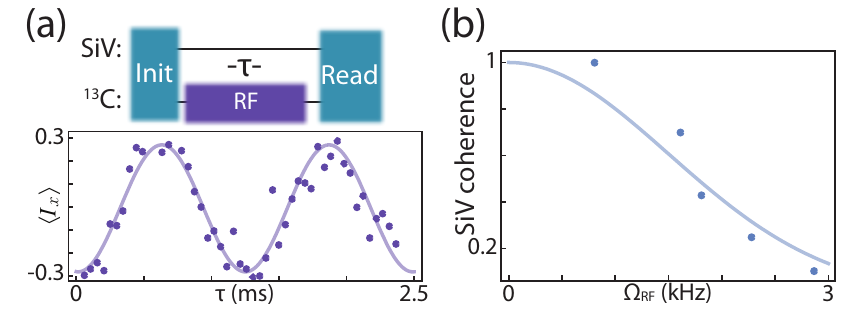}
	\caption{
			(a) RF Rabi oscillations. Applying an RF tone directly drives nuclear rotations of a coupled \cnuc.
			(b) SiV coherence in the presence of an RF drive. As the strength of the RF drive is increased, local heating  from the CPW reduces the SiV $T_2$.
			}
	    \label{fig:ngates}
    \end{figure}
		
The previous section demonstrated a CNOT gate between SiV and \cnuc\ using composite MW pulses. This approach has several drawbacks.
First, the gate fidelity is limited by our ability to finely tune the rotation angle of the maximally entangled gate which can not be done in a continuous fashion [see Fig.9(a)].
Second, this gate requires a specific number of MW pulses and delays between them, making the gate duration ($\sim$\SI{50}{\micro\second} in this work) comparable to the SiV coherence time.
Finally, this scheme relies on a second order splitting of individual \cnuc~resonances to resolve individual ones; residual coupling to additional \cnuc~limits the fidelity for a pulse sequence of given total length.

Direct RF control \cite{metsch2019initialization} would be a simple way to make a fast and high-fidelity CNOT gate since it would require a single RF $\pi$-pulse on a nuclear spin transition \cite{Rong2015experimental}.
Furthermore, since the nuclear spin transition frequencies depend on the hyperfine coupling to leading order, these pulses could have higher \cnuc~selectivity and potentially shorter gate duration.

We use the RF port inside the DR [Sec.~\ref{sec:mwcontrol}] to apply RF pulses resonant with nuclear spin transitions. 
Figure \ref{fig:ninit}(a) shows RF Rabi oscillations of the nuclear spin. 
Since the \cnuc\ gyromagnetic ratio is about 3 orders of magnitude smaller compared to the SiV spin, RF driving is much less efficient than MW one and requires much more power.
To investigate local heating of the SiV [Sec.~\ref{sec:mwcontrol}] we measured the SiV spin coherence contrast in spin-echo sequence right after applying off-resonant RF pulse of \SI{100}{\micro\second} at different power (calibrated via RF rabi oscillations) [Figure \ref{fig:ninit}(b)].
Unfortunately, Even modest Rabi frequencies ($\Omega_{\mathrm{RF}}\sim$\SI{1}{\kilo\hertz}) result in $20\%$ loss in SiV coherence.
Replacing the gold CWG used in this work by superconducting ones may solve heating issue and make RF driving practically useful. 

\section{conclusion}
The SiV center in diamond has rapidly become a leading candidate to serve as the building block of a future quantum network. 
In this work, we describe the underlying technical procedures and optimal parameter regimes necessary for utilizing the SiV-nanocavity system as a quantum network node. 
In particular, we discuss the effect of static and dynamic strain on the properties of the SiV spin qubit and its optical interface, with direct application to quantum networking experiments. 
We demonstrate techniques for coherently controlling and interfacing SiV spin qubits inside of nanophotonic structures at millikelvin temperatures to optical photons. 
Finally, we identify and coherently control auxiliary nuclear spins, forming a nanophotonic two-qubit register.

The work presented here and in the complementary letter \cite{nguyen2019quantum} illustrates the path towards the realization of a first-generation quantum repeater based on SiV centers inside diamond nanodevices. 
We note that a key ingredient enabling future, large-scale experiments involving several solid-state SiV-nanocavity nodes will be the incorporation of strain tuning onto each device \cite{machielse2019electromechanical}. 
Precise tuning of both the static and dynamic strain can overcome the limitations of inhomogeneous broadening and spectral diffusion, and enable scalable fabrication of quantum repeater nodes [Sec.~\ref{sec:strain}].


We would like to thank M. Markham, A. Bennett and D. Twitchen from Element Six Inc. for providing the diamond substrates used in this work, as well as F. Jelezko and R. Evans for insightful discussions. This work was supported by DURIP grant No. N00014-15-1-28461234 through ARO, NSF, CUA, AFOSR MURI, and ARL. M.K.B. and D.S.L were supported by DoD NDSEG, B.M. and E.N.K. were supported by NSF GRFP, and R.R. was supported by the Alexander von Humboldt Foundation. Devices were fabricated at Harvard CNS, NSF award no. 1541959.

\appendix

\section{Nanophotonic cavity design}
\label{apx:fab}

    \begin{figure}
		\includegraphics[width=\linewidth]{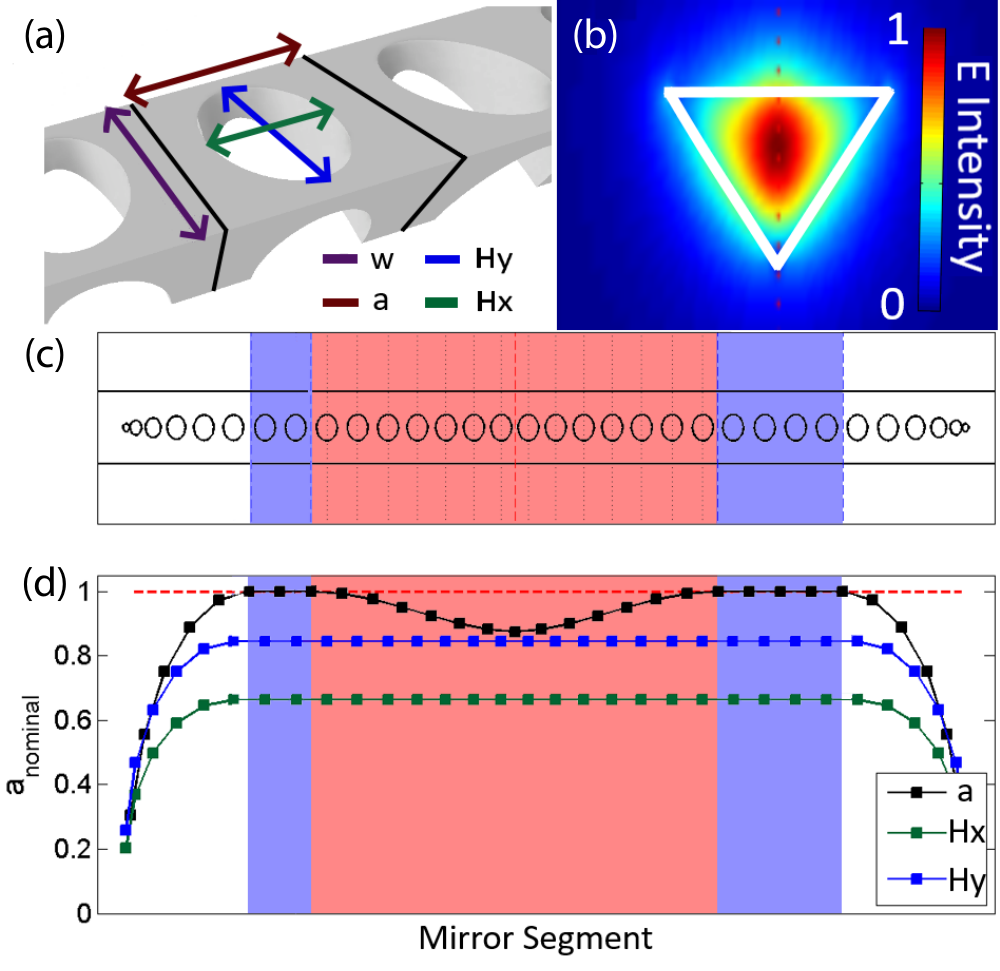}
	\caption{
			(a) Unit cell of a photonic crystal cavity (bounded by black lines). $H_x$ and $H_y$ define the size and aspect ratio of the hole, $a$ determines the lattice constant, and $w$ sets the waveguide width. 
			(b) Electric field intensity profile of the TE mode inside the cavity, indicating strong confinement of the optical mode inside the waveguide.
			(c) Schematic of photonic crystal design. Blue shaded region is the bandgap generating structure, red shaded region represents the cavity structure.
			(d) Plot of $a$, $H_x$, and $H_y$ for the cavity shown in (c), showing cubic taper which defines the cavity region. All sizes are shown in fractions of $a_{\mathrm{nominal}}$, the unperturbed lattice constant.
			}
	    \label{fig:simulation}
    \end{figure}
		
We simulate and optimize our nanophotonic structures to maximize atom-photon interactions while maintaining high waveguide coupling, which ensures good collection efficiency for the devices. 
In particular, this requires optimizing the device quality-factor to mode volume ratio, the relative rates of scattering into waveguide modes, and the size and shape of the optical mode. 
Each of these quantities are considered in a three-step simulation process (FDTD, Lumerical). 
We first perform a coarse parameter sweep over all possible unit cells which define the photonic crystal gemometry and identify families of bandgap-generating structures. 
These structures are the starting point for a gradient ascent optimization procedure, which results in generating high quality-factor, low mode volume resonators. 
Finally, the generated designs are modified to ensure efficient resonator-waveguide coupling.

Optimization begins by exploring the full parameter space of TE-like bandgap generating structures within our waveguide geometry. 
For hole-based cavities [Fig.~\ref{fig:simulation}(a)], this sweep covers a 5-dimensional parameter space: The lattice constant of the unit cell ($a$), the hole size and aspect ratio ($H_x$ and $H_y$), the device etch angle ($\theta$) and the waveguide width ($w$). 
Due to the size of this parameter space, we start by performing a low-resolution sweep over all parameters, with each potential design simulated by a single unit cell with the following boundary conditions: 4 perfectly matched layer (PML) boundary conditions in the transverse directions and 2 Bloch boundary conditions in the waveguide directions.
The band structure of candidate geometries are determined by sweeping the effective k-vector of the Bloch boundary condition and identifying allowed modes. 
Using this technique, families of similar structures with large bandgaps near the SiV transition frequency are chosen for further simulation. 
Each candidate photonic crystal is also inspected for the position of its optical mode maximum, ensuring that it has first-order modes concentrated in the center of the diamond, where SiVs will be incorporated [Fig.~\ref{fig:simulation}(b)].

The second step is to simulate the full photonic crystal cavity design, focused in the regions of parameter space identified in step one. This is done by selecting a fixed $\theta$, as well as a total number of unit cells that define the structure, then modifying the bandgap of the photonic crystal with a defect region to form a cavity mode. 
We define this defect using a cubic tapering of one (or several) possible parameters:
\begin{equation}
A(x) = 1-d_{\mathrm{max}}|2x^3 -3x^2+1|
\end{equation}
where $A$ is the relative scale of the target parameter(s) at a distance $x$ from the cavity center, and $d_{\mathrm{max}}$ is the defect depth parameter. 
Photonic Crystal cavities with multi-parameter defects are difficult to reliably fabricate, therefore, devices used in this work have cavity defect geometries defined only by variations in the lattice constant.
The cavity generated by this defect is scored by simulating the optical spectrum and mode profile and computing the scoring function $F$:
\begin{equation}
F = \mathrm{min}(Q,Q_{\mathrm{cutoff}})/(Q_{\mathrm{cutoff}}\times V_{\mathrm{mode}})
\end{equation}
Where $Q$ is the cavity quality-factor, $Q_{\mathrm{cutoff}} = 5\times 10^5$ is an estimated maximum realizable $Q$ based on fabrication constraints, and $V_{\mathrm{mode}}$ is the cavity mode volume. 
Based on this criteria, we employ a gradient ascent process over all cavity design parameters (except $\theta$ and the total number of unit cells) until $F$ is maximized, or a maximum number of iterations has occurred. 
Due to the complexity and size of the parameter space, a single iteration of this gradient ascent is unlikely to find the optimal structure. 
Instead, several candidates from each family of designs found in step one are explored, with the best moving on to the final step of the simulation process. 
These surviving candidates are again checked to ensure confinement of the optical mode in the center of the cavity structure and to ensure that the structures fall within the tolerances of the fabrication process.

The final step in the simulation process is to modify the optimized designs to maximize resonator-waveguide coupling. 
This is done by removing unit cells from the input port of the device, which decreases the overall quality-factor of the devices in exchange for better waveguide damping of the optical field. 
Devices are once again simulated and analyzed for the fraction of light leaving the resonator through the waveguide compared to the fraction scattering into free-space. 
The number of unit cells on the input port is then optimized for this ratio, with simulations indicating that more than $95\%$ of light is collected into the waveguide. 
In practice, fabrication defects increase the free-space scattering rate, placing resonators close to the critically-coupled regime. 
Finally, the waveguide coupling fraction is increased by appending a quadratic taper to both ends of the devices such that the optical mode is transferred adiabatically from the photonic crystal region into the diamond waveguide. 
This process produces the final cavity structure used for fabrication [Fig.~\ref{fig:simulation}(c)].

\section{Strain-induced frequency fluctuations}
\label{apx:strain}
In this Appendix we calculate changes the SiV spin-qubit frequency and optical transition frequency arising from strain fluctuations.
We start with the Hamiltonian for SiV in an external magnetic field $B_z$ aligned along trhe SiV symmetry axis \cite{hepp2014electronic, meesala2018strain}:
\begin{multline}
H=-\lambda 
\underbrace{
	\begin{pmatrix}
		0 & 0	&	i	&	0 	\\
		0 & 0	&	0	&	-i	\\
	 -i	&	0	&	0	&	0		\\
		0	&	i	&	0	&	0
	\end{pmatrix} 
}_\text{spin-orbit}
+
\underbrace{
	\begin{pmatrix}
		\alpha - \beta & 0	&	\gamma	&	0 	\\
		0 & \alpha - \beta	&	0	&	\gamma	\\
	 \gamma	&	0	&	\beta	&	0		\\
		0	&	\gamma	&	0	&	\beta
	\end{pmatrix}
}_\text{strain}
+\\
\underbrace{
	q \gamma_L B_z
	\begin{pmatrix}
		0 & 0	&	i	&	0 	\\
		0 & 0	&	0	&	i	\\
	 -i	&	0	&	0	&	0		\\
		0	&	-i	&	0	&	0
	\end{pmatrix}
}_\text{orbital Zeeman}
+
\underbrace{
	\frac{\gamma_S B_z}{2}
	\begin{pmatrix}
		1 & 0	&	0	&	0 	\\
		0 & -1	&	0	&	0	\\
	  0	&	0	&	1	&	0		\\
		0	&	0	&	0	&	-1
	\end{pmatrix}
}_\text{spin Zeeman},
\label{eqn:apxH}
\end{multline}
where $\lambda$ is a spin-orbit coupling constant, $\gamma_L=\mu_B$ and $\gamma_S=2\mu_B$ are Land\'{e} g-factors of the orbital and spin degrees of freedom ($\mu_B$ the Bohr magneton), $q=0.1$ is a Ham reduction factor of the orbital momentum \cite{slichter1990magnetic, hepp2014electronic}, and $\alpha, \beta, \gamma$ are local strain parameters which can be different for the ground and excited sates [Sec.~\ref{sec:strain}]. 
As measuring the exact strain parameters is challenging [Sec.~\ref{sec:strain}] we assume only one non-zero component in this tensor ($\epsilon_{zx}$) in order to simplify our calculations.
In this case, strain parameters are:
\begin{align}
	\beta &= f_\text{g(e)}\epsilon_{zx},\\
	\alpha &= \gamma = 0,
\end{align}
where $f_\text{g(e)}=1.7\times 10^6\, (3.4 \times 10^6)$\,GHz/strain \cite{meesala2018strain} for the ground (excited) state and the GS splitting is:
\begin{equation}
	\Delta_{GS} = 2 \sqrt{\lambda_g^2+\beta^2},
\end{equation}
where $\lambda_g \approx 25$\,GHz is the SO-constant for the ground state.
Next, we solve this Hamiltonian and investigate how the qubit frequency changes as a function of relative strain fluctuations ($\xi$):
\begin{equation}
\Delta f_\text{MW} = 
\frac{2 \left(f_\text{g}\epsilon_{zx} \right)^2 \lambda_g B_z q \gamma_L}
{{\left( \left(f_\text{g}\epsilon_{zx} \right)^2+\lambda_g^2 \right)}^{3/2}}
\xi.
\end{equation}
The corresponding change in the optical frequency is:
\begin{equation}
\Delta f_\text{optical} = 
	\left(
		\frac{\left(f_\text{g}\epsilon_{zx}\right)^2}
				 {\sqrt{\left(f_\text{g}\epsilon_{zx}\right)^2+\lambda_g^2}}
		-
		\frac{\left(f_\text{e}\epsilon_{zx}\right)^2}
				 {\sqrt{\left(f_\text{e}\epsilon_{zx}\right)^2+\lambda_e^2}}
	\right)
	\xi,
\end{equation}
where 
$\lambda_e \approx 125$\,GHz is the SO-constant for the excited state.

For SiV 2 [Sec.~\ref{sec:strain}] we measured $\Delta_{GS}= 140$\,GHz and find $\epsilon_{zx} = 3.8\times  10^{-5}$. 
With $\xi = 1\%$ strain fluctuations (corresponding to $\sim 10^{-7}$ strain), frequencies change by $\Delta f_\text{MW} \approx 4$\,MHz and $\Delta f_\text{optical} \approx -300$\,MHz.
This quantitatively  agrees with the data presented in [Fig.~\ref{fig:sivstrain}(f)]. 

\section{Mitigating spectral diffusion}
\label{apx:diffusion}
    \begin{figure}
		\includegraphics[width=\linewidth]{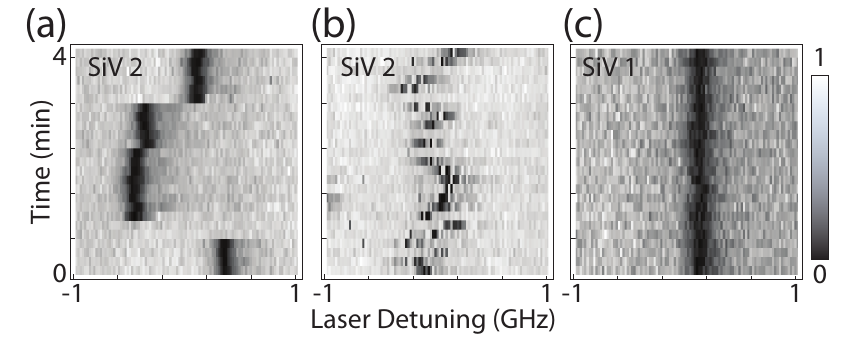}
	\caption{ 
			(a) Spectral diffusion of SiV 2. We observe slow spectral wandering as well as spectral jumps. 
			(b) Applying a short green repumping pulse before every measurement significanly speeds up the timescale for spectral diffusion.
			(c) Spectral diffusion of SiV 1 in nanostructures. Line is stable to below \SI{100}{\mega\hertz} over many minutes. 
			Scale bar indicates normalized SiV reflection signal.
						}
	    \label{fig:diffusion}
    \end{figure}
In order to couple SiV centers to a quantum network, electronic transitions must be stabilized with respect to a probe laser. 
We note that such spectral diffusion is a universal challenge for solid-state quantum systems \cite{faraon2012coupling, Kuhlmann2015transform, neill2013Fluctuations}. 
In the case of the SiV center, spectral diffusion can be seen explicitly in figure \ref{fig:diffusion}(a), where the optical transition frequency can either drift slowly (central region), or undergo large spectral jumps. 
As this diffusion can be larger than the SiV linewidth, any given instance of an experiment could have the probe laser completely detuned from the atomic transition, resulting in a failed experiment. 

There are several possible solutions to mitigate this spectral diffusion.
First, exploiting a high-cooperativity interface, one can Purcell-broaden the optical linewidth [sec.~\ref{sec:cavity}] to exceed the spectral diffusion \cite{sipahigil2016integrated}.
Second, a high collection efficiency can be used to read out the optical position faster than the spectral diffusion. 
The frequency can then be probabilistically stabilized by applying a short laser pulse at \SI{520}{\nano\meter} which dramatically speeds up the timescale of spectral diffusion, \cite{bhaskar2017quantum, evans2018photon} [Fig.~\ref{fig:diffusion}(b)].
Alternatively this signal could be used to actively stabilize the line using strain-tuning \cite{Sohn2018controlling, machielse2019electromechanical}. From the observations in figure \ref{fig:sivstrain}(f), this technique should mitigate spectral diffusion of both the optical and spin transitions. Strain tuning also offers the capability to control the DC strain value, which has important effects on qubit properties as discussed previously, and enables tuning multiple SiV centers to a common network operation frequency. As such, this tunability will likely be an important part of future quantum networking technologies based on SiV centers.

The severity of spectral diffusion is different for different emitters however, and this control is not always necessary, especially for proof-of-principle experiments with a small number of emitters. 
For SiV 1, the main SiV used in the following sections, and the SiV used in ref \cite{nguyen2019quantum}, we find almost no spectral diffusion, with optical transitions stable over many minutes [Fig.~\ref{fig:diffusion}(c)]. 
This is an ideal configuration, as experiments can be performed without any need to verify the optical line position. 

\section{Model for SiV decoherence}
\label{apx:bath}
The scaling of $T_2(N) \propto N^{2/3}$ is identical to that found for nitrogen-vacancy centers, where it is assumed that $T_2$ is limited by a fluctuating electron spin bath \cite{deLange2010Universal, Myers2014Probing}. 
Motivated by DEER measurements with SiV 2, we follow the analysis of ref. \cite{Myers2014Probing} to estimate the noise bath observed by SiV 1. 

The measured coherence decay is modeled by:
\begin{equation}
\langle S_z\rangle = \mathrm{Exp}\left(-\int \mathrm{d}\omega\, S(\omega)\mathcal{F}_N(t, \omega)\right), 
\end{equation} 
where $S(\omega)$ is the noise power-spectrum of the bath, and $\mathcal{F}_N(t, \omega) = 2\sin(\omega t/2)(1-\sec(\omega t/2N))^2/\omega^2$ is filter function for a dynamical-decoupling sequence with an even number of pulses \cite{Myers2014Probing}. 
We fit sucessive $T_2$ echo curves to the functional form $A+Be^{-(t/T_2)^\beta}$, with $A, B$ being free parameters associated with photon count rates, and $\beta=3$ providing the best fit to the data. 
This value of $\beta$ implies a decoherence bath with a Lorentzian noise power-spectrum, $S(\omega, b, \tau) = b^2\tau/\pi \times 1/(1+\omega^2\tau^2)$, where $b$ is a parameter corresponding to the strength of the noise bath, and $\tau$ is a parameter corresponding to the correlation time of the noise \cite{deLange2010Universal, Myers2014Probing}.

Empirically, no one set of noise parameters faithfully reproduces the data for all measured echo sequences. 
Adding a second source of dephasing $\tilde{S} = S(\omega, b_1, \tau_1) + S(\omega, b_2, \tau_2)$, gives reasonable agreement with the data using parameters $b_1 = $\SI{5}{\kilo\hertz}, $\tau_1 = $\SI{1}{\micro\second}, $b_2 = $\SI{180}{\kilo\hertz}, $\tau_2 = $\SI{1}{\milli\second} [Fig.~\ref{fig:mwdeer}(d)]. 
The two drastically different set of noise parameters for each of the sources can help illuminate the source of noise in our devices. 

As explained in the previous section, one likely candidate for this decoherence is a bath of free electrons arising from improper surface termination or local damage caused during nanofabrication, which are known to have correlation times in the $\sim$\si{\micro\second} range. 
The SiV studied in this analysis is approximately equidistant from three surfaces: the two nearest holes which define the nanophotonic cavity, and the top surface of the nanobeam [sec:~\ref{sec:fab}], all of which are approximately \SI{50}{\nano\meter} away. We estimate a density of $\sigma_{\mathrm{surf}} = 0.067$ spins/\si{\nano\meter^2} using:
\begin{equation}
b_1 = \gamma_{\mathrm{SiV}}\langle B_{\mathrm{surf}}\rangle = \frac{\mathrm{g}^2\mu_B^2\mu_0}{\hbar}\frac{1}{4\pi \Sigma d_i^2}\sqrt{\frac{\pi}{4\sigma_{\mathrm{surf}}}}
\label{eq:noisestr}
\end{equation}
where $b_1$ is the measured strength of the noise bath, $\mathrm{g}$ is the electron gyromagnetic ratio, and $d_i$ are the distances to the nearest surfaces. This observation is consistent with surface spin densities measured using NVs \cite{Myers2014Probing}. 

The longer correlation time for the second noise term suggests a different bath, possibly arising from free electron spins inside the bulk diamond. 
Vacancy clusters, which can persist under annealing even at \SI{1200}{C}, are known to posses $g=2$ electron spins, and are one possible candidate for this noise bath \cite{Yamamoto2013extending}. 
Integrating over $d$ in eq. \ref{eq:noisestr}, we estimate the density of spins required to achieve the measured $b_2$. We estimate $\rho_\mathrm{bulk}\sim 0.53$ spins per \si{\nano\meter^3}, which corresponds to a doping of $3$ppm. 
Interestingly, this is nearly identical to the local concentration of silicon incorporated during implantation (most of which is not successfully converted into negatively charged SiV), and could imply implantation-related damage as a possible source of these impurities.  

Another possible explanation for this slower bath could be coupling to nuclear spins in the environment. The diamond used in this experiment has a natural abundance of \cnuc, a spin-1/2 isotope, in concentrations of approximately $1.1\%$. Replacing $\mu_B \rightarrow \mu_N$ in the term for $\langle B\rangle$ gives an estimated nuclear spin density of $\rho_{\mathrm{bulk,N}} = 0.6\%$, only a factor of two different than the expected nuclear spin density.

\section{Concurrence and Fidelity calculations}
\label{apx:fid}

\subsection{Spin-photon concurrence and fidelity calculations}

From correlations in the Z- and X-bases, we estimate a lower bound for the entanglement in our system. Following reference \cite{Chou2005measurement}, we note that the density matrix of our system conditioned on the detection of one photon can be described as: 

\begin{equation}
\rho_{ZZ} = 1/2
\begin{pmatrix}
p_{e\uparrow}& 0& 0& 0\\
0& p_{e\downarrow}& c_{e\downarrow, l\uparrow}& 0\\
0& c_{e\downarrow, l\uparrow}^\dagger& p_{l\uparrow}& 0\\
0& 0& 0& p_{l\downarrow}
\end{pmatrix}
\label{eq:rho}
\end{equation}
where $p_{ij}$ are the probabilities of measuring a photon in state $i$, and the spin in state $j$. 
$c_{e\downarrow, l\uparrow}$ represents entanglement between $p_{e\uparrow}$ and $p_{l\downarrow}$. 
We set all other coherence terms to zero, as they represent negligibly small errors in our system (for example, $c_{e\uparrow, e\downarrow}>0$ would imply that the SiV was not initialized properly at the start of the measurement). 
We quantify the degree of entanglement in the system by its concurrence $\mathcal{C}$, which is $0$ for seperable states, and $1$ for a maximally entangled state \cite{Wootters1998Entanglement}: 
\begin{equation}
\mathcal{C} = \mathrm{Max}(0,\lambda_0^{1/2} - \sum_{i=1}^N \lambda_i^{1/2}),
\end{equation}
where $\lambda_i$ are the eigenvalues of the matrix $\rho_{ZZ} \cdot \left(\sigma_y \cdot \rho_{ZZ} \cdot \sigma_y^\dagger\right)$, and $\sigma_y$ is the standard Pauli matrix acting on each qubit basis separately ($\sigma_y = \sigma_{y,\mathrm{ph}}\otimes\sigma_{y,\mathrm{el}}$). 
While this can be solved exactly, the resulting equation is complicated. Taking only the first-order terms, this can be simplified to put a lower bound on the concurrence: 
\begin{equation}
\mathcal{C} \geq 2(|c_{e\downarrow, l\uparrow}| - \sqrt{p_{e\uparrow}p_{l\downarrow}})
\end{equation}
We measure $p$ direcly in the Z basis, and estimate $|c_{e\downarrow, l\uparrow}|$ by performing measurements in the X basis. 
A $\pi/2$-rotation on both the photon and spin qubits rotates:

\begin{align*}
|e\rangle &\rightarrow 1/\sqrt2(|e\rangle + |l\rangle),\quad |l\rangle \rightarrow 1/\sqrt2(|e\rangle - |l\rangle)\\
|\downarrow\rangle &\rightarrow 1/\sqrt2(|\downarrow\rangle + |\uparrow\rangle),\quad |\uparrow\rangle \rightarrow 1/\sqrt2(|\downarrow\rangle - |\uparrow\rangle)
\end{align*}

Afer this transformation, the signal contrast directly measures $c_{e\downarrow, l\uparrow}$:

\begin{equation}
2 c_{e\downarrow, l\uparrow} = p_{-,\leftarrow} + p_{+,\rightarrow} - p_{-\rightarrow} - p_{+\leftarrow}\\
\Rightarrow \mathcal{C} \geq 0.42(6)
\end{equation}

Similarly, the fidelity of the entangled state (post-selected on the detection of a photon) can be computed by the overlap with the target Bell state \cite{bernien2013heralded}: 

\begin{equation}
F = \langle \Psi^+|\rho_{ZZ}|\Psi^+\rangle = (p_{e\uparrow} + p_{l\downarrow} + 2 c_{e\downarrow, l\uparrow})/2 \geq 0.70(3)
\end{equation}

\subsection{Correcting for readout infidelity}

Errors arising from single-shot readout incorrectly assign the spin state, results in lower-contrast histograms for spin-photon correlations. 
We follow the analysis done in ref.~\cite{bernien2013heralded}, and correct for readout errors using a transfer matrix formalism. The measured spin-photon correlations $p_{ij}$ are related to the `true' populations $P_{ij}$ via:

\begin{equation}
\begin{pmatrix}
p_{e\downarrow}\\
p_{e\uparrow}\\
p_{l\downarrow}\\
p_{l\uparrow}
\end{pmatrix}
=
\begin{pmatrix}
F_{\downarrow}& 1-F_{\uparrow}& 0& 0\\
1-F_{\downarrow}& F_{\uparrow}& 0& 0\\
0& 0& F_{\downarrow}& 1-F_{\uparrow}\\
0& 0& 1-F_{\downarrow}& F_{\uparrow}
\end{pmatrix}
\begin{pmatrix}
P_{e\downarrow}\\
P_{e\uparrow}\\
P_{l\downarrow}\\
P_{l\uparrow}
\end{pmatrix}
\end{equation}

with $F_{\downarrow},\quad F_{\uparrow}$ defined above. 
After this correction, an identical analysis is performed to calculate the error-corrected histograms [Fig.~\ref{fig:spinphoton}(b,c,d) dark-shading]. We find an error-corrected concurrence $\mathcal{C} \geq 0.79(7)$ and fidelity $F \geq 0.89(3)$.

\subsection{Electron-nuclear concurrence and fidelity calculations}
For spin-spin Bell states, in contrast to the spin-photon analysis, we can no longer set any of the off-diagonal terms of the density matrix [eq.~\ref{eq:rho}] to zero due to the limited ($\sim 90\%$) nuclear initialization fidelity. 
We note that neglecting these off-diagonal terms can only decrease the estimated entanglement in the system, thus the concurrence can still be written as:

\begin{equation}
\mathcal{C} \geq 2(|c_{\downarrow\uparrow}| - \sqrt{p_{\uparrow\uparrow}p_{\downarrow\downarrow}})
\end{equation}
where the first subscript is the electron spin state, and the second is the nuclear state. 
We estimate $c_{\downarrow\uparrow}$ again by using the measured populations in an orthogonal basis. In this case, off-diagonal terms add a correction:

\begin{equation}
2 c_{\downarrow\uparrow} + 2c_{\uparrow\downarrow} = p_{\leftarrow\leftarrow} + p_{\rightarrow\rightarrow} - p_{\leftarrow\rightarrow} - p_{\rightarrow\leftarrow}
\end{equation}

In order for the density matrix to be properly normalized, $c_{\uparrow\downarrow} \leq \sqrt{p_{\uparrow\uparrow}p_{\downarrow\downarrow}}$, giving us the final concurrence:

\begin{equation}
\mathcal{C} \geq p_{\leftarrow\leftarrow} + p_{\rightarrow\rightarrow} - p_{\leftarrow\rightarrow} - p_{\rightarrow\leftarrow} - 4\sqrt{p_{\uparrow\uparrow}p_{\downarrow\downarrow}}
\end{equation}

Additionally, both electron readout error as well as \cnuc\ mapping infidelity can misreport the true spin state. As such, the new transfer matrix to correct for this error is:
\onecolumngrid
\begin{equation}
\begin{pmatrix}
F_{\downarrow,e}F_{\downarrow,N}& F_{\downarrow,e}(1-F_{\uparrow,N})& (1-F_{\uparrow,e})F_{\downarrow,N}& (1-F_{\uparrow,e})(1-F_{\uparrow,N})\\
F_{\downarrow,e}(1-F_{\downarrow,N})& F_{\downarrow,e}F_{\uparrow,N}& (1-F_{\uparrow,e})(1-F_{\downarrow,N})& (1-F_{\uparrow,e})F_{\uparrow,N}\\
(1-F_{\downarrow,e})F_{\downarrow,N}& (1-F_{\downarrow,e})(1-F_{\uparrow,N})& F_{\uparrow,e}F_{\downarrow,N}& F_{\uparrow,e}(1-F_{\uparrow,N})\\
(1-F_{\downarrow,e})(1-F_{\downarrow,N})& (1-F_{\downarrow,e})F_{\uparrow,N}& F_{\uparrow,e}(1-F_{\downarrow,N})& F_{\uparrow,e}F_{\uparrow,N}
\end{pmatrix}
\end{equation}
\twocolumngrid

Where $F_{\downarrow,e}\approx F_{\uparrow,e} = 0.85$ and $F_{\downarrow,N}\approx F_{\uparrow,N} = 0.72$. 
Following this analysis, we report an error-corrected concurrence of $\mathcal{C} \geq 0.22(9)$.

\subsection{Electron-nuclear CNOT gate}
We further characterize the CNOT gate itself as a universal quantum gate. 
Due to the relatively poor readout fidelity (see above), we do not do this by performing quantum state tomography. 
Instead, we estimate entries in the CNOT matrix using measurements in only the Z-basis. 
As a control measurement, we first initialize the two qubits in all possible configurations and read out, averaged over many trials. 
Next, we initialize the qubits, perform a CNOT gate, and read out, again averaged over many trials, normalized by the control data. 
Any reduction in contrast after normalization is attributed to the opposite spin state, establishing a system of equations for determining the CNOT matrix. 
We solve this system of equations, marginalizing over free parameters to determine a MLE estimate for the CNOT transfer matrix, as seen in reference \cite{nguyen2019quantum}.

\section{Nuclear initialization and readout}
\label{apx:init}

\begin{figure}
	\includegraphics[width=\linewidth]{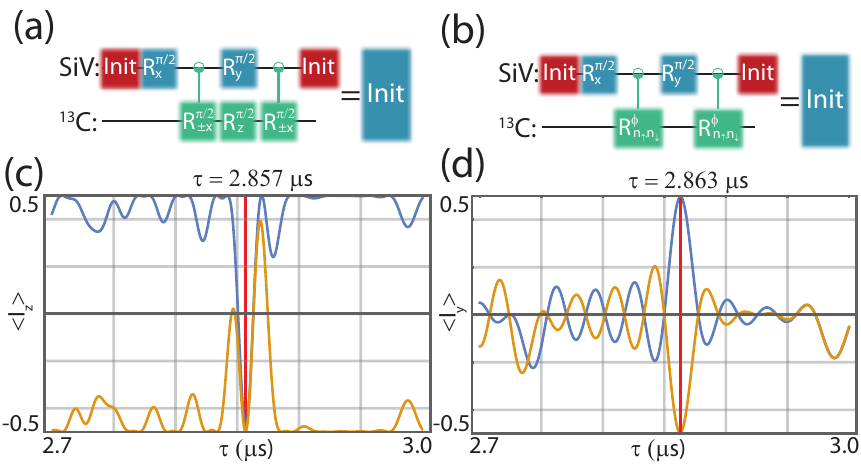}
	\caption{
			(a) Original initialization sequence from~\cite{taminiau2014universal}, 
			note $\mathcal{R}_{z,\mathrm{C}}^{\pi / 2}$ rotation.
			(b) Simplified initialization sequence used in this work.
			(c) Simulated performance of the initialization gate from (b) 
			using 8 $\pi-$pulses per each nuclear gate, the initial state is 
			$\left|\uparrow \uparrow \right\rangle$ (blue) and 
			$\left|\uparrow \downarrow \right\rangle$ (orange). The resonances 
			are narrow compared to (d) due to applying effectively 
			twice more $\pi-$pulses
			(d) Simulated performance of $\mathcal{R}_{\pm x,\mathrm{SiV-C}}^{\pi / 2}$ 
			gate for 8 $\pi-$pulses for SiV-\cnuc\ register initialized in 
			$\left|\uparrow \uparrow \right\rangle$ (blue) and 
			$\left|\downarrow \uparrow \right\rangle$ (orange)
		}
	  \label{fig:AppendB}
\end{figure}
		
Initialization (and readout) of the \cnuc\ spin can be done by mapping population between the SiV spin and the \cnuc. 
Following reference \cite{taminiau2014universal}, we note that Z and X gates are possible with dynamical-decoupling based nuclear gates, thus a natural choice for initialization are gates comprised of both $\mathcal{R}_{\pm x,\mathrm{SiV-C}}^{\pi / 2}$ and $\mathcal{R}_{z,\mathrm{SiV-C}}^{\pi / 2}$, as shown in figure \ref{fig:AppendB}(a) and in reference \cite{taminiau2014universal}.
We note here that it should be possible to combine the effects of $\mathcal{R}_{x}$ and $\mathcal{R}_{z}$ rotations in a single gate, which has the potential of shortening and simplifying the total initialzation gate. 
One proposed sequence uses the following entangling gate:
\begin{multline}
\mathcal{R}_{\vec{n_\uparrow},\vec{n_\downarrow}}^{\phi}=
	\begin{pmatrix}
		(1+i)/2	&	i/\sqrt{2}	&	0	&	0\\
		i/\sqrt{2}	&	(1-i)/\sqrt{2}	&	0	&	0\\
		0	&	0	&	(1+i)/2	&	-i/\sqrt{2}\\
		0	&	0	&	-i/\sqrt{2}	&	(1-i)/2
	 \end{pmatrix}\\
	=\begin{pmatrix}
		R_{\Theta = \pi/4}^{\pi/2}\,R_{z}^{\pi/2} &	0\\
		0	&	R_{\Theta = \pi/4}^{-\pi/2}\,R_{z}^{\pi/2}
	\end{pmatrix} 
\end{multline}
which corresponds to a rotation on the angle $\phi=2\pi /3$ around the axes $n_{\uparrow, \downarrow}=\{\pm \sqrt{2},0,1\}/\sqrt{3}$.
The matrix of entire initialization gate [Fig.~\ref{fig:AppendB}(b)] built from this gate would then be:
\begin{equation}
\text{Init}=
	\begin{pmatrix}
		0 & 0	&	-(1+i)/2	&	-1/\sqrt{2} 	\\
		i/\sqrt{2} & -(1+i)/2	&	0	&	0	\\
	  0	&	0	&	-(1-i)/2	&	-i/\sqrt{2}		\\
		1/\sqrt{2} &	(1-i)/2	&	0	&	0
	\end{pmatrix}
\end{equation}
which results in an initialized \cnuc\ spin.

To demonstrate this, we numerically simulate a MW pulse sequence using the exact coupling parameters of our \cnuc~\cite{nguyen2019quantum} and 8 $\pi-$pulses for each $\mathcal{R}_{\vec{n_\uparrow},\vec{n_\downarrow}}^{\phi}$ gate.
Figure~\ref{fig:AppendB}(c) shows that regardless of the initial state, the \cnuc\ always ends up in state $\left|\downarrow \right\rangle$ (given that the SiV was initialized in $\left|\uparrow \right\rangle$).
As expected, the timing of this gate ($\tau_\text{init} =$ \SI{2.857}{\micro\second}) is noticeably different from the timing of the $\mathcal{R}_{\pm x,\mathrm{SiV-C}}^{\pi / 2}$ gate ($\tau_{\pi / 2} =$ \SI{2.851}{\micro\second}), which occurs at spin-echo resonances [Fig.~\ref{fig:AppendB}(d)].

The rotation matrix for this sequence at $\tau=\tau_\text{init}$ (with the SiV initialized in $\left|\uparrow \right\rangle$) is:
\begin{equation}
\mathcal{R}_{n_\uparrow}^{\phi}=
	\begin{pmatrix}
		0.55 +0.51 i & 0+0.65 i	\\
		0.65 i & 0.55-0.52 i
	 \end{pmatrix}
\end{equation}
corresponding to a rotation angle $\phi = 0.63\pi$ around the axis $n_\uparrow=\{0.78,0,0.62\}$, very close to the theoretical
result.

Since the experimental fidelities for both initialization gates [Fig.~\ref{fig:AppendB} (a,b)] are similar, we use sequence (b) to make the gate shorter and avoid unnecessary pulse-errors.

\bibliography{SiVbib}
\end{document}